\documentclass{emulateapj}

\newcommand{\brg}{Br${\gamma}$ }

\begin{document}

\title{Measuring Distance and Properties of the Milky Way's Central 
Supermassive Black Hole with Stellar Orbits}

\author{
A. M. Ghez\altaffilmark{1,2},
S. Salim\altaffilmark{1,4},
N. N. Weinberg\altaffilmark{3,5},
J. R. Lu\altaffilmark{1},
T. Do\altaffilmark{1},
J. K. Dunn\altaffilmark{1},
K. Matthews\altaffilmark{3},
M. Morris\altaffilmark{1},
S. Yelda\altaffilmark{1},
E. E. Becklin\altaffilmark{1},
T. Kremenek\altaffilmark{1},
M. Milosavljevic\altaffilmark{6},
J. Naiman\altaffilmark{1,7}
}

\altaffiltext{1}{UCLA Department of Physics and Astronomy, Los Angeles, CA 90095
-1547; ghez, jlu, tdo, jkdunn, morris, syelda, becklin@astro.ucla.edu}
\altaffiltext{2}{UCLA Institute of Geophysics and Planetary Physics,
Los Angeles, CA 90095-1565}
\altaffiltext{3}{California Institute of Technology, Division of Mathematics, Physics and Astronomy, Pasadena, CA 91125; kym@caltech.edu}
\altaffiltext{4}{NOAO, 950 N Cherry Ave, Tucson, AZ 85719, samir@noao.edu}
\altaffiltext{5}{University of California Berkeley, Department of Astronomy\\
Berkeley, CA 94720-3411 nnw@astron.berkeley.edu}
\altaffiltext{6}{University of Texas, Department of Astronomy, Austin, TX 78712 milos@astro.as.utexas.edu}
\altaffiltext{7}{UCSC, Department of Astronomy \& Astrophysics, Santa Cruz, CA 95064, jnaiman@astro.ucsc.edu}

\begin{abstract}
We report new precision measurements of the properties of our Galaxy's 
supermassive black hole. Based on astrometric (1995-2007) and radial 
velocity (2000-2007) measurements from the W. M. Keck 10-meter telescopes, a 
fully unconstrained Keplerian orbit for the short period star S0-2 provides 
values for the distance (R$_0$) of 8.0 $\pm$  0.6 kpc, the enclosed mass 
(M$_{bh}$) of 4.1 $\pm$ 0.6 $\times$ $10^6 M_{\odot}$, and the black hole's 
radial velocity, which is consistent with zero with 30 km/s uncertainty. 
If the black hole is assumed to be at rest with respect to the Galaxy 
(e.g., has no massive companion to induce motion), we can further constrain 
the fit and obtain R$_0$ = 8.4 $\pm$ 0.4 kpc 
and M$_{bh}$ = 4.5 $\pm$ 0.4 $\times$ $10^6 M_{\odot}$. 
More complex models constrain the extended dark mass distribution to be less 
than 3-4 $\times$ $10^5 M_{\odot}$ within 0.01 pc, $\sim$100x higher than 
predictions from stellar and stellar remnant models.  
For all models, we identify transient astrometric shifts from source 
confusion (up to 5x the astrometric error) and the assumptions regarding the 
black hole's radial motion as previously unrecognized limitations on orbital 
accuracy and the usefulness of fainter stars.  Future astrometric and RV 
observations will remedy these effects.  
Our estimates of R$_0$ and the Galaxy's local rotation speed, which 
it is derived from combining R$_0$ with the apparent proper motion of Sgr A*,
($\theta_0$ = 229 $\pm$ 18 km s$^{-1}$),
are compatible with measurements made using other methods. 
The increased black hole
mass found in this study, compared to that determined using projected mass 
estimators, implies a longer period for the innermost stable orbit, longer 
resonant relaxation timescales for stars in the vicinity of the black hole
and a better agreement with the M$_{bh}$-$\sigma$ relation.

\end{abstract}

\keywords{black hole physics -- Galaxy:center --- 
Galaxy:kinematics and dynamics ---
infrared:stars -- techniques:high angular resolution}

\section{INTRODUCTION}
\label{sec:intro}

Ever since the discovery of fast moving (v $>$ 1000 km s$^{-1}$) stars within
0.$\tt''$3 (0.01 pc) of our Galaxy's central supermassive black hole 
(Eckart \& Genzel 1997; Ghez et al. 1998), the prospect of using stellar orbits 
to make precision measurements of the black hole's mass (M$_{bh}$) and
kinematics, the distance to the Galactic center (R$_0$) and, more ambitiously, 
to measure post-Newtonian effects has been anticipated
(Jaroszynski 1998, 1999; Salim \& Gould 1999; Fragile \& Mathews 2000;
Rubilar \& Eckart 2001; Weinberg, Milosavlejic \& Ghez 2005; Zucker
\& Alexander 2007; Kraniotis 2007; Will 2008).  
An accurate measurement of the Galaxy's central black hole 
mass is useful for putting the Milky Way in context with other
galaxies through the apparent 
relationship between the mass of the central black hole
 and the 
velocity dispersion, $\sigma$, of the host galaxy
(e.g., Ferrarese  \& Merrit 2000; Gebhardt et al. 2000; Tremaine et al. 2002).
It can also be used as a test of this scaling, as the
Milky Way has the most convincing case for a supermassive 
black hole of any galaxy used to define this relationship.
Accurate estimates of R$_0$ impact a wide range of issues associated with the 
mass and structure of the Milky Way, including possible constraints on the 
shape of the dark matter halo and
the possibility that the Milky Way is a lopsided spiral (e.g., Reid 1993; Olling \& Merrifield 2000;
Majewski et al. 2006).  Furthermore, if measured with sufficient
accuracy ($\sim$1\%), the distance to the Galactic center could influence the 
calibration
of standard candles, such as RR Lyrae stars, Cepheid variables and giants, used in
establishing the extragalactic distance scale. 
In addition to estimates of $M_{bh}$ and R$_0$, precision measurements of 
stellar kinematics offer the exciting possibility of detecting
deviations from a Keplerian orbit.
This would allow an exploration of a possible 
cluster of stellar remnants surrounding
the central black hole, suggested by Morris (1993), Miralda-Escud{\'e} \& Gould(2000), and Freitag et al. (2006).
Estimates for the mass of the remnant cluster
range from $10^4 - 10^5 M_{\odot}$ within a few tenths of a parsec of the central black hole.  Absence of such a remnant cluster 
would be interesting
in view of the hypothesis that the inspiral of intermediate-mass black holes
by dynamical friction could deplete any centrally concentrated cluster
of remnants.  Likewise, measurements of post-newtonian effects would 
provide a test of general relativity, and, ultimately, could probe the spin of the central 
black hole. 

Tremendous observational progress has been made over the last decade towards 
obtaining accurate estimates of the orbital parameters of the fast
moving stars at the Galactic center.  Patience alone permitted new 
astrometric measurements that yielded the first accelerations (Ghez et al. 2000; 
Eckart et al. 2002), which suggested that the orbital period of 
the best characterized star, S0-2, could be as short as 15 years.  The passage of more time then led to 
full astrometric orbital solutions (Sch\"odel et al. 2002, 2003; Ghez et al. 2003, 2005a),
which increased the implied dark mass densities by a factor of $10^4$ 
compared to earlier velocity dispersion work and thereby solidified the case for a supermassive 
black hole.  The advent of adaptive optics enabled radial velocity measurements of these stars (Ghez et al. 2003), which permitted the first 
estimates of the distance to the Galactic center from stellar orbits
(Eisenhauer et al. 2003, 2005).  

In this paper, we present new orbital models for S0-2.  These provide the 
first estimates of the distance to the Galactic center and limits on the 
extended mass distribution
based on data collected with the W. M. Keck telescopes.
The ability to probe the properties of the Galaxy's central supermassive 
black hole has benefitted from several advancments since our previous
report (Ghez et al. 2005).  First,
new astrometric and radial velocity 
measurements have been collected between 2004 and 2007, increasing the 
quantity of kinematic data available.  Second,  the majority of the new data 
was obtained with the laser guide star adaptive optics system at Keck, 
improving 
the quality of the measurements (Ghez et al. 2005b; Hornstein et al. 2007).
These new data sets are presented in \S\ref{sec:obs}.
Lastly, new data analysis has improved our ability to extract 
radial velocity estimates from past
spectroscopic measurements, allowing us to extend 
the radial velocity curve back in time by two years, as described 
in \S\ref{sec:data_analysis}.  The orbital analysis, described in \S\ref{sec:orbit}, 
identifies several sources of previously unrecognized biases 
and the implications of
our results are discussed in \S\ref{sec:disc}.

\section{OBSERVATIONS \& DATA SETS}
\label{sec:obs}

\subsection{High Angular Resolution Imaging: Speckle and Adaptive Optics}

For the first eleven years of this experiment (1995-2005), 
the proper motions of stars orbiting the center of our Galaxy
were obtained from $K$[2.2 $\mu m$]-band speckle observations 
of the central stellar cluster with the W. M. Keck I 10-meter telescope and
its facility near-infrared camera, NIRC (Matthews \& Soifer 1994;
Matthews et al.\ 1996).  
A total of 27 epochs of speckle observations 
are included in the analysis conducted in this paper, of which 22 have been 
reported in earlier papers by our group (Ghez et al. 1998, 2000, 2005a).
Five new speckle observations, between 2004 April and 2005 June, were 
conducted in a similar manner.
In summary, during each observing run, 
$\sim$10,000 short ($t_{exp}$ = 0.1 sec) exposure frames
were obtained with NIRC in its fine plate scale mode, which has a scale of
20.46 $\pm$ 0.01  mas pixel$^{-1}$ (see Appendix B) and a corresponding field 
of view of 
5\farcs 2 $\times$ 5\farcs 2.  Interleaved with these observations were 
similar sequences on a dark patch of sky.  From these data, we produce 
images that are diffraction-limited ($\theta$ = 0\farcs 05) and have
Strehl ratios of $\sim$0.05.
 
With the advent of laser guide star adaptive optics (LGSAO) in 2004 on the
10 m W. M. Keck II telescope (Wizinowich et al. 2006; van Dam et al. 2006), 
we have made measurements of the Galaxy's central stellar cluster with much 
higher Strehl ratios (Ghez et al. 2005b).
Between 2004 and 2007, nine LGSAO data sets were taken
using the W. M. Keck II facility near-infrared camera, NIRC2 (P.I.
K. Matthews), which has an average plate scale of 9.963 $\pm$ 0.006 mas 
pixel$^{-1}$ (see Appendix C) 
and a field of view of 10\farcs 2 $\times$ 10\farcs 2.  All but one of the
observations were obtained through a
K' ($\lambda_0$=2.12 $\mu$m, $\Delta \lambda$=0.35 $\mu$m) band-pass filter, with
the remaining one obtained through narrow band filters 
(CO: $\lambda_0$ = 2.278 $\mu$m, $\Delta \lambda$ = 0.048 $\mu$m
and
Kcont: $\lambda_0$ = 2.27 $\mu$m, $\Delta \lambda$ = 0.030 $\mu$m).
During these observations, the laser guide star's position was fixed to
the center of the camera's field of view and therefore moved when the telescope
was dithered.  While the laser guide star is used to 
correct most of the important atmospheric aberrations,
it does not provide information on the tip-tilt term, which, for all our
LGSAO observations (imaging and spectroscopy),
was obtained from visible observations of USNO 0600-28577051
(R = 13.7 mag and $\Delta r_{SgrA*}$ = 19$\arcsec$).
Details of the observing setup for 2004 July 26, 2005 June 30, and 2005 July 31 
are described in detail in Ghez et al. (2005b), Lu et al. (2008), and 
Hornstein et al. (2007), respectively.  While each of these early LGSAO 
observations had a slightly different setup and dither pattern, the more 
recent, deeper, LGSAO measurements (2006-2007) were obtained with nearly 
identical setups.  
Specifically, we used a 20 position dither pattern with 
randomly distributed (but repeatable) positions in a
0\farcs 7 $\times$ 0\farcs 7 box and an initial position that placed
IRS 16NE on pixel (229, 720) at a sky PA set to 0.0.  This setup keeps 
the brightest star in the region, IRS 7 (K=6.4), off the 
field of view at all times.
At each position, three exposures, each composed of 10 coadded 2.8 sec 
integrations,
were obtained; the integration time was set 
with the aim of keeping the detector's response linear 
beyond the full width at half maximum (FWHM) point for the 
brightest (K=9.0) star in the field of view;
the number of images 
per position was chosen to provide the minimum elapsed time 
needed to allow the LGSAO system's optimization algorithm to converge 
($\sim$3 min.) before dithering.  Table \ref{tbl_img} summarizes all the new 
imaging data sets.

\begin{deluxetable*}{llrrllllll}
\tabletypesize{\scriptsize}
\tablewidth{0pt}
\tablecaption{Summary of New Keck Imaging Observations\label{tbl_img}}
\tablehead{
	\colhead{Date (UT)} &
	\colhead{Technique\tablenotemark{a}} &
	\colhead{Frames} &
	\colhead{Frames} &
	\colhead{Coadd $\times$ T$_{exp}$} & 
	\colhead{FWHM} &
	\colhead{Strehl} &
	\colhead{Number} &
	\colhead{K$_{lim}$\tablenotemark{c}\tablenotemark{d}} &
	\colhead{Pos. Error\tablenotemark{e}} \\
	\colhead{ } &
	\colhead{ } & 
	\colhead{Obtained} & 
	\colhead{Used} &
	\colhead{(sec)} &
	\colhead{(mas)} &
	\colhead{ } &
	\colhead{of Stars\tablenotemark{b}\tablenotemark{d}} &
	\colhead{mag} &
	\colhead{mas} 
}
\startdata
2004 April 29-30               & Speckle  & 20140 & 1444 & 1 $\times$ 0.137  & 63 & 0.09 & 163 & 15.9 & 0.9  \\ 
2004 July 25-26                & Speckle  & 14440 & 2156 & 1 $\times$ 0.137  & 61 & 0.07 & 165 & 15.9 & 0.9  \\
2004 August 29	               & Speckle  & 3040  & 1300 & 1 $\times$ 0.137  & 60 & 0.08 & 138 & 15.7 & 1.0  \\
2005 April 24-25               & Speckle  & 15770 & 1677 & 1 $\times$ 0.137  & 60 & 0.07 & 143 & 15.6 & 0.9  \\
2005 July 26-27                & Speckle  & 14820 & 1825 & 1 $\times$ 0.137  & 62 & 0.05 & 116 & 15.5 & 1.2  \\
2004 July 26	               & LGSAO(1) & 12    &   12 & 50 $\times$ 0.181 & 60 & 0.31 & 233 & 16.0 & 0.3  \\
2005 June 30  			& LGSAO(2) & 10	  &    10 & 5 $\times$ 7.2/11.9\tablenotemark{f}    & 61 & 0.32 & 269 & 16.4 & 1.3 \\
2005 July 30-31	               & LGSAO(3) &  66	  &   32 & 10 $\times$ 2.8   & 62 & 0.34 & 565 & 19.0 & 0.19 \\
2006 May 3	               & LGSAO(4) & 153	  &  107 & 10 $\times$ 2.8   & 58 & 0.30 & 562 & 19.2 & 0.16 \\
2006 June 20-21                & LGSAO(4) & 295   &  152 & 10 $\times$ 2.8   & 57 & 0.33 & 580 & 19.1 & 0.10 \\
2006 July 17	               & LGSAO(4) &  70   &   64 & 10 $\times$ 2.8   & 59 & 0.31 & 574 & 19.2 & 0.19 \\
2007 May 17                    & LGSAO(4) & 103   &   77 & 10 $\times$ 2.8   & 58 & 0.35 & 566 & 19.1 & 0.21 \\
2007 May 20	               & LGSAO(1) & 20	  &   12 & 10 $\times$ 2.8   & 77 & 0.20 & 394 & 17.8 & 0.28 \\
2007 Aug 10, 12                & LGSAO(4) & 142	  &   79 & 10 $\times$ 2.8   & 57 & 0.32 & 553 & 19.1 & 0.20 \\
\enddata
\tablenotetext{a}{For the LGSAO data sets, the number in parentheses denotes
the observational setup used (e.g., dither pattern and camera orientation; 
see \S2.1 for details).}
\tablenotetext{b}{The number of stars detected within 3 arcsec of SgrA*.}
\tablenotetext{c}{K$_{lim}$ is the
magnitude at which the cummulative distribtuion function of the observed K magnitudes
reaches 90\% of the total sample size.}
\tablenotetext{d}{For this analysis
only stars in 4 or more epochs are considered to eliminate any spurious
source detections.}
\tablenotetext{e}{The average positional uncertainty due to centroiding in each epoch is estimated
from a set of 25 stars detected in all epcohs and brighter than K$\sim$13 mag.}
\tablenotetext{f}{ Half of the images were taken using a narrow band CO filter,
with the shorter exposure time,
and the other half using a narrow band K$_{cont}$ filter, with the 
longer exposure time.}
\end{deluxetable*}


\subsection{Adaptive Optics Spectroscopy}

To monitor the line-of-sight motions of stars orbiting the center of our Galaxy
between the years 2000 and 2007, 
high angular resolution spectroscopic
observations of stars in the Sgr A* stellar cluster
were taken with both the natural guide star adaptive optics (NGSAO;
Wizinowich et al. 2000)
system (2000-2004) and the LGSAO system (2005-2007) on
the W. M. Keck II 10 m telescope.  
The NGSAO atmospheric corrections and the LGSAO tip-tilt corrections 
were made on the basis of
visible observations of USNO 0600-28579500 
(R = 13.2 mag and $\Delta r$ $\sim$ 30$\arcsec$)
and USNO 0600-28577051
(R = 13.7 mag and $\Delta r$ $\sim$ 19$\arcsec$), respectively.
While the angular resolution of the NGSAO spectra was typically 2-3 times
the diffraction limit ($\theta_{diff}$ = 54 mas), a point spread function (PSF) FWHM 
of $\sim$ 70 mas at 2 $\micron$ was achieved for the LGSAO long exposure 
spectra. 

Three different spectrometers have been used over the course of this study.
Our earliest measurements were obtained in 2000 June with
NIRSPEC (McLean et al. 1998, 2000) 
in its low resolution slit spectrometer mode (R $\sim$ 2600).  It was not 
originally designed to go behind the adaptive optics system and therefore
had inefficient throughput in its AO mode; it was, however, the only
spectrometer available behind the AO system in 2000.  While the resulting
low signal to noise data set yielded no line detections in the initial analysis 
of S0-2 (Gezari et al. 2002), we now have the advantage of knowing
what type of lines are present in the spectra
and have therefore included
this data set in our analysis by retroactively identifying the Br$\gamma$ line, 
which is used to measure radial velocities (see \S3.2)
Between 2002 and 2005, NIRC2 (P.I. K. Matthews) was used in
its spectroscopic R $\sim$ 4000 mode, which is generated with 
a 20 mas pixel scale, a medium-resolution 
grism and a 2 pixel slit.  In 2002, this 
produced the first line detection in S0-2 (Ghez et al. 2003) and, since then,
three new NIRC2 measurements (2 with NGSAO and 1 with LGSAO) have been 
obtained.  Since 2005, OSIRIS, which is an integral field spectrograph
with a 2 $\micron$ spectral resolution of $\sim$ 3600
(Larkin et al. 2006), has been used.  The field of view of 
this spectrograph depends on the pixel scale and filter.
Most of the OSIRIS observations were taken using the
35 mas pixel scale and the narrow band filter Kn3 (2.121 to 2.229 $\micron$; 
includes Br$\gamma$), which results in a field of view of 
$1.\arcsec 12 \times 2.\arcsec 24$, and were centered on S0-2.
All of the OSIRIS observations were obtained with the LGSAO system. 
Table \ref{tbl_spec} summarizes the details of the 
10 new spectroscopic measurements of S0-2 that were made between the years
2003 and 2007 (see Gezari et al 2002 \& Ghez et al. 2003 for details of the
2000-2002 measurements).  

\begin{deluxetable*}{lllllll}
\tabletypesize{\scriptsize}
\tablewidth{0pt}
\tablecaption{Summary of New Keck Spectroscopic Observations\label{tbl_spec}}
\tablehead{
	\colhead{Date} &
	\colhead{Instrument/} &
	\colhead{Filter: Spectral} & 
	\colhead{Pix.} &
	\colhead{Num. Exp. } &
	\colhead{SNR\tablenotemark{b}} &
	\colhead{Calibration Stars}\\
	\colhead{(UT)} &
	\colhead{AO System} &
	\colhead{Range ($\mu$m)} &
	\colhead{Scale } &
	\colhead{$\times$ T$_{exp}$ (sec)} &
	\colhead{} &
	\colhead{(G2/A0)} \\
	\colhead{} &
	\colhead{} &
	\colhead{} &
	\colhead{(mas)} &
	\colhead{} &
	\colhead{} &
	\colhead{} 
}
\startdata
%
%
%
2003 June 08    & NIRC2\tablenotemark{a} / NGS     & K: 2.08 - 2.34 	& 20 &  2 $\times$ 1200 & 62 & HD 193193 / HD 195500 \\
2004 June 22    & NIRC2\tablenotemark{a} / NGS     & K': 2.00 - 2.26 	& 20 & 16 $\times$ 1200 & 23 & HD 193193 / HD 195500 \\
2005 May 30    & NIRC2\tablenotemark{a} / LGS  	 & K': 2.00 - 2.26 	& 20 &  7 $\times$ 1200 & 31 & HD 198099 / HD 195500 \\
%
2005 July 03    & OSIRIS / LGS	 & Kbb: 1.97 - 2.39 	& 20 &  7 $\times$  900 & 30 & HD 193193 / HD 195500 \\
2006 May 23    & OSIRIS / LGS	 & Kbb: 1.97 - 2.39   	& 35 &  4 $\times$  900 & 25 & HD 193193 / HD 195500  \\  
2006 June 18   & OSIRIS / LGS	 & Kn3: 2.121 - 2.229 	& 35 & 10 $\times$  900 & 52 & HD 198099 / HD 195500  \\ 
2006 June 30   & OSIRIS / LGS	 & Kn3: 2.121 - 2.229 	& 35 &  9 $\times$  900 & 33 & HD 193193 / HD 195500  \\
2006 July 1    & OSIRIS / LGS    & Kn3: 2.121 - 2.229 	& 35 &  9 $\times$  900 & 60 & HD 150437 / HD 155379  \\
2007 May 21     & OSIRIS / LGS    & Kn3: 2.121 - 2.229 	& 35 &  3 $\times$  900 & 28 & HD 198099 / HD 195500  \\
2007 July 18-19	& OSIRIS/LGS	 & Kn3: 2.121 - 2.229   & 35 &  2 $\times$  900 & 22  & HD 193193 / HD 195500 \\ 
\enddata
\tablenotetext{a}{For the NIRC2 data sets, the slit position angle was
259.4$^o$ (2003), 333.76$^o$ (2004), and 355.9$^o$ (2005).}
\tablenotetext{b}{The SNR is per spectral pixel and is calculated 
between 2.13 and 2.145 $\mu$m.  The width of a spectral pixel 
is roughly 2.5 and 2.53 \AA ~ for OSIRIS and NIRC2, respectively.}

\end{deluxetable*}

\section{DATA EXTRACTION}
\label{sec:data_analysis}

\subsection{Image Analysis \& Astrometry}

\begin{figure*}
\epsscale{1.0}
\plotone{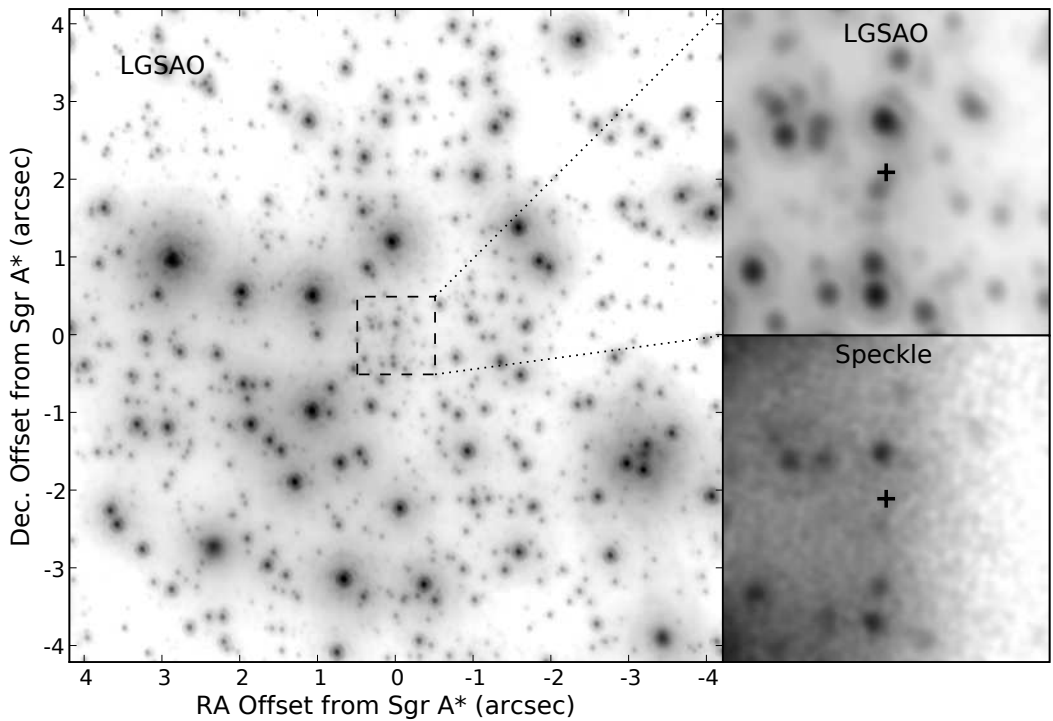}
\figcaption{
A comparison of raw images obtained with LGSAO and speckle imaging with the 
Keck 10 m telescopes.  The large scale image is an LGSAO image obtained in
2005.  The inset LGSAO image (top right) and speckle image (bottom right) 
are centered on the black hole, SgrA*
(marked with a cross), 
with a field of view of 1\farcs 0 $\times$ 
1\farcs 0, also obtained in 2005.
The image quality, depth, and astrometric precision
have all been greatly improved with the advent of LGSAO.
\label{fig_aoVsp}
}
\end{figure*}

The individual speckle and adaptive optics data frames are processed in 
two steps to create a final average image for each of the 34 imaging 
observing runs.  First, each frame is sky-subtracted, flat fielded, 
bad-pixel-corrected, corrected for distortion effects
and, in the case of the speckle data, resampled by a factor of two;  
the distortion
correction applied to the NIRC2/LGSAO data is from the NIRC2 pre-ship review 
results 
(\url{http://www2.keck.hawaii.edu/inst/nirc2/preship\_testing.pdf}) 
and those
applied to the speckle data sets are the combined transformations given
in Ghez et al. (1998) and Lu et al. (2008). 
The frames are then registered on the basis of the position of IRS 16C, 
for the speckle images, and a crosss-correlation of the entire image, for the 
LGSAO image,  
and combined.  For the adaptive optics data sets, the frames whose
PSF has a FWHM $<$ 1.25 x FWHM$_{min}$, where FWHM$_{min}$ is the 
minimum observed FWHM for each epoch and which typically includes $\sim$70\% of 
the measured frames,
are combined with a weighted average with weights set equal to their strehl 
ratios.
To increase the signal to noise ratio of the 2005 June data set, 
the data taken through the two narrow-band filters are averaged together.
For the speckle data set, only the best $\sim$ 2,000 frames from each 
observing run are combined using a weighted ``Shift-and-Add" technique 
described by Hornstein (2007).  
The selected frames from each observing run (speckle and LGSAO) are also divided
into three independent subsets from which three subset images are created
in a similar manner to the average images; these subset images are used to 
assess photometric and  astrometric measurement uncertainties.  
Figure \ref{fig_aoVsp} shows examples of the final average LGSAO and speckle 
images.  While all the images sets have point spread function (PSF) 
cores that are nearly diffraction-limited 
($\theta$ $\sim$ 0\farcs 06 vs. $\theta_{diff.~lim}$ = 0\farcs05), 
the LGSAO images have much higher image quality 
than the speckle images, 
with median Strehl ratios of $\sim$0.3 and 0.07, for the LGSAO and speckle
images, respectively. 

\begin{figure*}
\epsscale{1.0}
\plotone{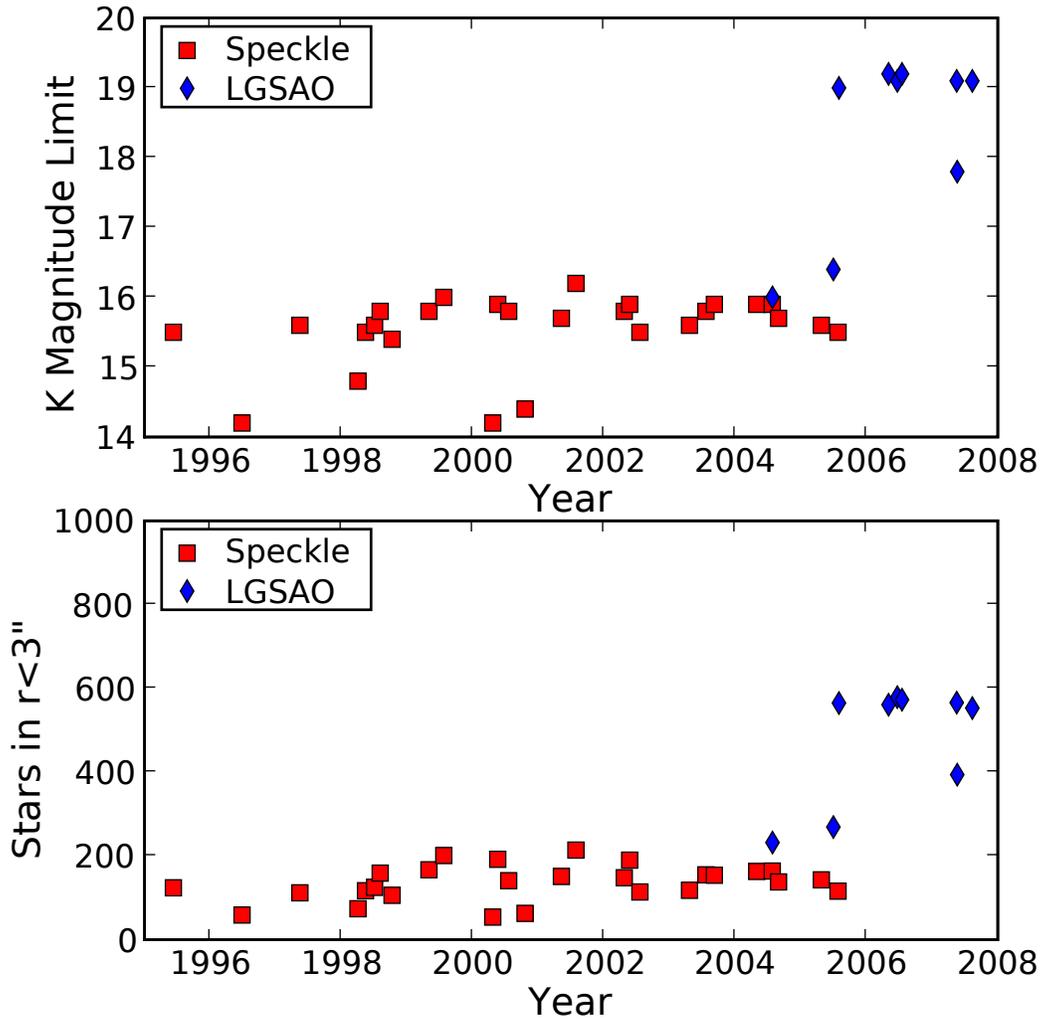}
\figcaption{
Comparison of the sensitivity of the average images from each epoch.
The recent LGSAO images, with significantly longer on-sky integration times
(t$_{tot}$ $\sim$ 50 vs.\ 3 min) and much higher strehl ratios,
are three magnitudes more sensitive than any of the speckle images.
\label{speck_ao_klimit}
}
\end{figure*}

\begin{figure*}
\epsscale{1.0}
\plotone{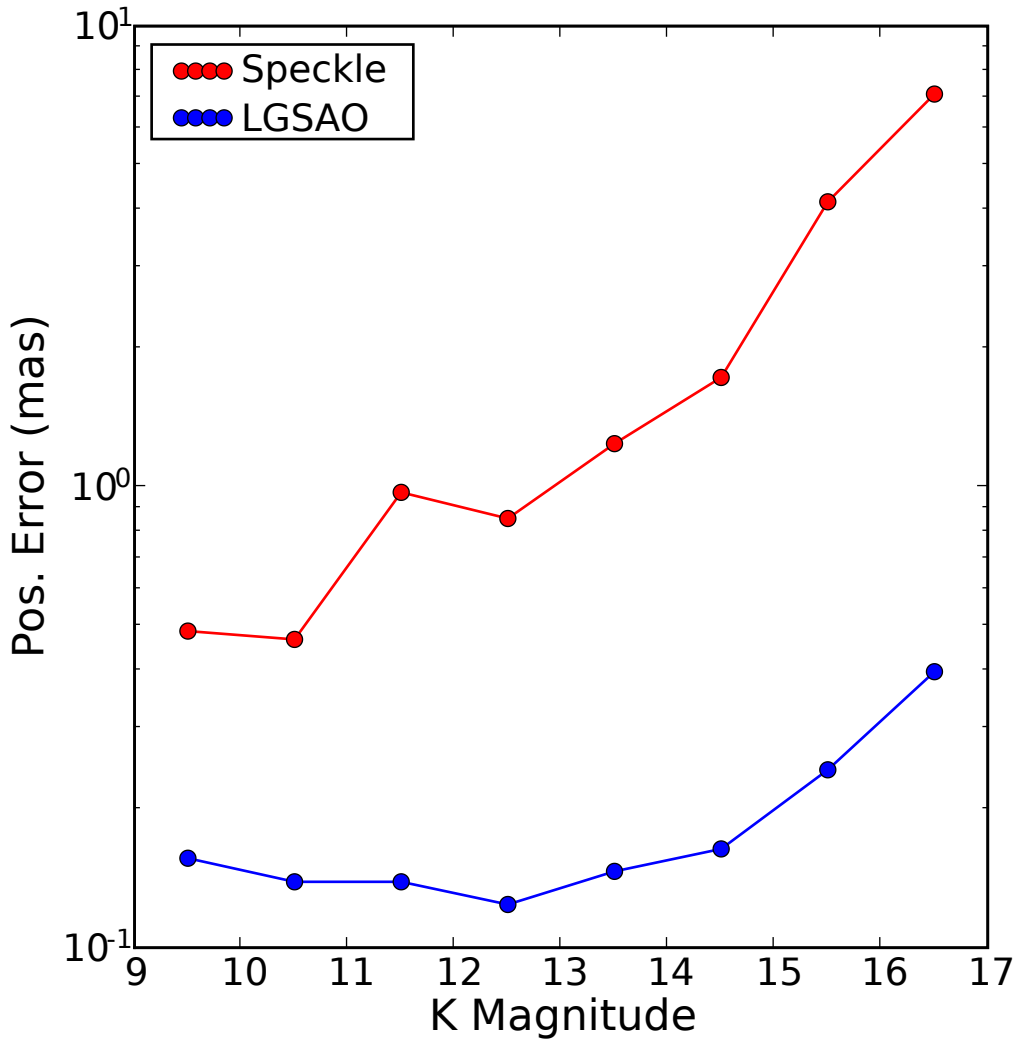}
\figcaption{
Comparison of the centroid uncertainties as a function of brightness.
Because the very brightest stars (K$\sim$9) are saturated in their
cores in the LGSAO images, there is a slight rise in their centroid
uncertainties compared to somewhat fainter sources.  Overall, however,
for bright sources (K$<$13), the long exposure LGSAO images achieve a centroiding uncertainty of just 0.17 mas, 
a factor of $\sim$6 better than the earlier work done with 
speckle imaging.  
\label{speck_ao_astro}
}
\end{figure*}

Point sources are identified and characterized in each of the images
using the PSF fitting program 
StarFinder (Diolaiti et al. 2000) on both the average images and the subset
images.  StarFinder iteratively generates a PSF
based on user selected point sources\footnote 
{In this analysis, the stars that are input into the PSF construction
are IRS 16C, 16NW, and S2-17 for the speckle images and IRS 16C, 16NW,
16NE, 16SW, 33E, 33W, 7, 29N, and GEN+2.33+4.60 for the LGSAO images.}
in the image and identifies
additional sources in the image by cross-correlating the resulting PSF with the 
image.
The initial source list for each image is composed only 
of sources detected in the average images with correlation values
above 0.8 and in all three subset images with correlation values above 0.6.
Eleven bright (K$<$14 mag), non-variable sources establish 
the photometric zero points for each list based on measurements made
by Rafelski et al. (2007; IRS 16C, IRS 16SW-E,
S2-17, S1-23, S1-3, S1-4, S2-22, S2-5, S1-68, S0-13, S1-25).
As shown in Figure \ref{speck_ao_klimit}, the deep LGSAO images 
(K$_{lim} \sim $19 mag) are 
three magnitudes more sensitive than the speckle images 
(K$_{lim} \sim $16 mag),
which results in roughly three times 
more sources being detected in the LGSAO images than the speckle images over a 
comparable region.  
Because of the higher signal to noise, as shown in Figure \ref{speck_ao_astro}, the centroiding 
uncertainties ($\delta X', \delta Y'$),
which  are estimated from the RMS error of the measurements in 
the three subset images, are a factor of 6 more precise 
for the deep LGSAO data sets (0.17 mas) than the speckle data sets (1.1 mas),
for bright stars (K$<$13 mag); the plateau observed in the relative centroiding 
uncertainties for the brighter stars (K$<$13) in the LGSAO images is 
likely caused by the combined effects of 
differential tip-tilt jitter 
and residual optical distortions across the field of view.

The sources identified each night are matched across multiple epochs
and their positions are transformed to a common coordinate system that 
will be referred to as the {\it cluster reference frame}.  As detailed in
Appendix A, the transformation for each epoch is derived by minimizing
the net displacement of a set of 
``coordinate reference" stars, allowing for proper motions, 
relative to their positions in 
a common reference image, which, in this case, is the 2004 July LGSAO 
image.  This procedure attempts to ensure that 
in the cluster reference frame the 
coordinate reference stars are at rest (i.e., 
no net translation, rotation, expansion,
or skew). 
A total of $\sim$470 and $\sim$120 stars serve as coordinate reference
stars in the LGSAO and speckle epochs, respectively. 
These stars are selected based on the following criteria:
(1) high detection correlations ($>$0.9), ensuring good positional 
accuracy, (2) located more than 0\farcs 5 from Sgr A* 
to avoid sources with measurable 
non-linear motions 
(i.e., accelerations in the plane of the sky $> \sim$8 km/s/yr), 
(3) low velocities ($<$ 15 mas/yr, or equivalently $\sim$600 km s$^{-1}$), which 
eliminates
possible coordinate reference sources that have been mismatched across epochs,
and 
(4) lack of spectroscopic identification as a young star 
from Paumard et al. (2006) to eliminate the known net rotation of the
young stars in the cluster reference frame. 
Positional uncertainties from this transformation process, which are 
characterized by a half sample bootstrap applied to the coordinate
reference stars, are    
a factor of $\sim$1.5 (speckle) to 6 (LGSAO) smaller than the centroiding 
uncertainties and grow by less than a factor of 2 between the center of 
the field of view (minimum) and a radius of 3\arcsec.  

An additional source of positional error originates from 
residual optical distortion in NIRC2.  While the residual distortion
in NIRC2 is small, the extremely precise centroid measurements in the deep 
LGSAO images make it a significant effect.
The presence of such a systematic error
is established by examining the distribution of positional
residuals, normalized by measurement (centroiding plus alignment) 
uncertainties, 
to the linear proper motion fits
for the coordinate reference stars.
The speckle data sets do not show large, measurable 
biases;  
the speckle measurements, on average, are only  
1$\sigma$ off from the linear proper motion fit.
In contrast, the much more precise deep-LGSAO astrometric 
measurements are, on average, 5$\sigma$ off from these fits. 
As described in Appendix B, we account for this effect
at two stages of our analysis.  First, 0.88 mas is added in quadrature
to the positional uncertainties of the coordinate reference stars to account 
for systematic errors in the coordinate transformations.  Second, a local correction, 
in the coordinate reference frame, is derived and applied to the positions of 
the short period stars that were made with LGSAO setups that differ from that 
of the reference image.  This procedure ensures that residuals from
both linear proper motion fits to the coordinate reference stars 
(see Appendix A \& B) and from orbit fits to S0-2 (see \S4) are 
consistent with a normal distribution.

\begin{figure*}
\epsscale{1.0}
\plotone{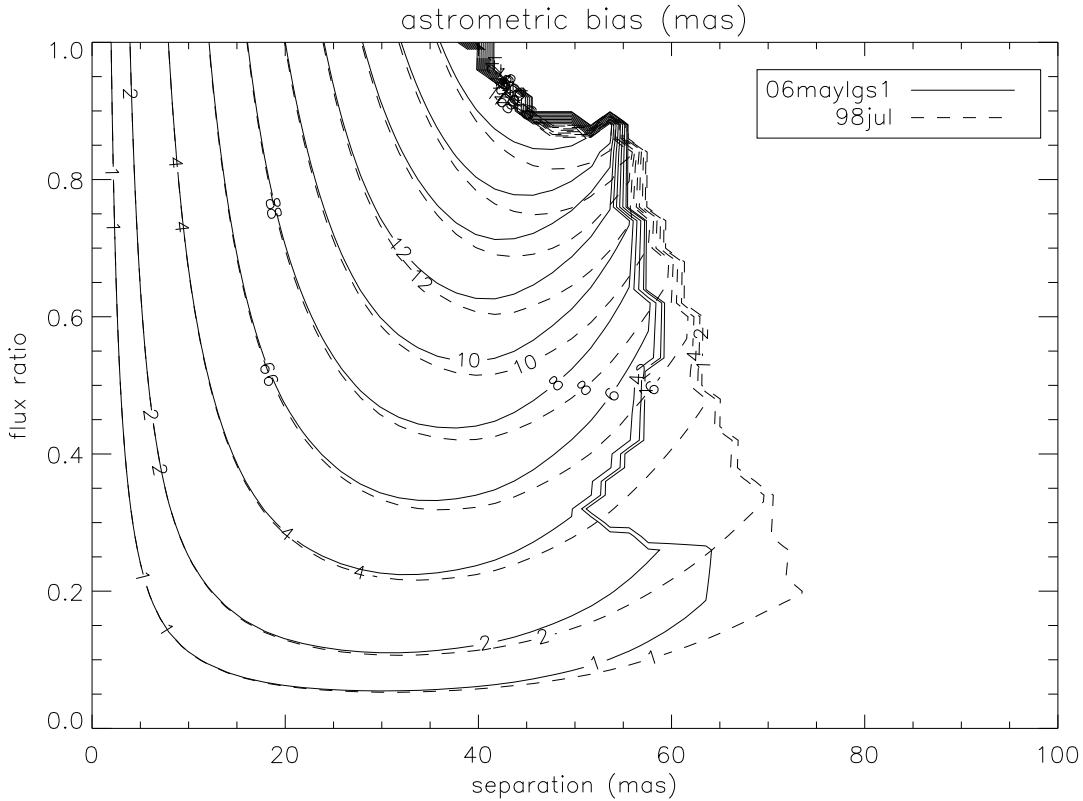}
\figcaption{
Astrometric bias introduced by an unresolved source in the case
of a binary star generated and analyzed with a known PSF.  Two cases are
shown: (solid line) PSF from LGSAO image in 2006 May and (dotted line) PSF from
speckle image in 1998 July.  The contour lines show the amount
of bias (in mas) introduced by an underlying source of the
indicated flux ratio and separation.  Once the neighboring source is detected,
which happens at separations of $\sim$60 mas, the astrometric bias drops to 
zero in this idealized case.  For 1:1 binaries, pairs with smaller 
separations can be resolved.  This figure shows that biases well above 
the positional uncertainties ($\sim$1 mas) can occur due to underlying sources. 
\label{bias_binary}
}
\end{figure*}

\begin{figure*}
\epsscale{1.0}
\plotone{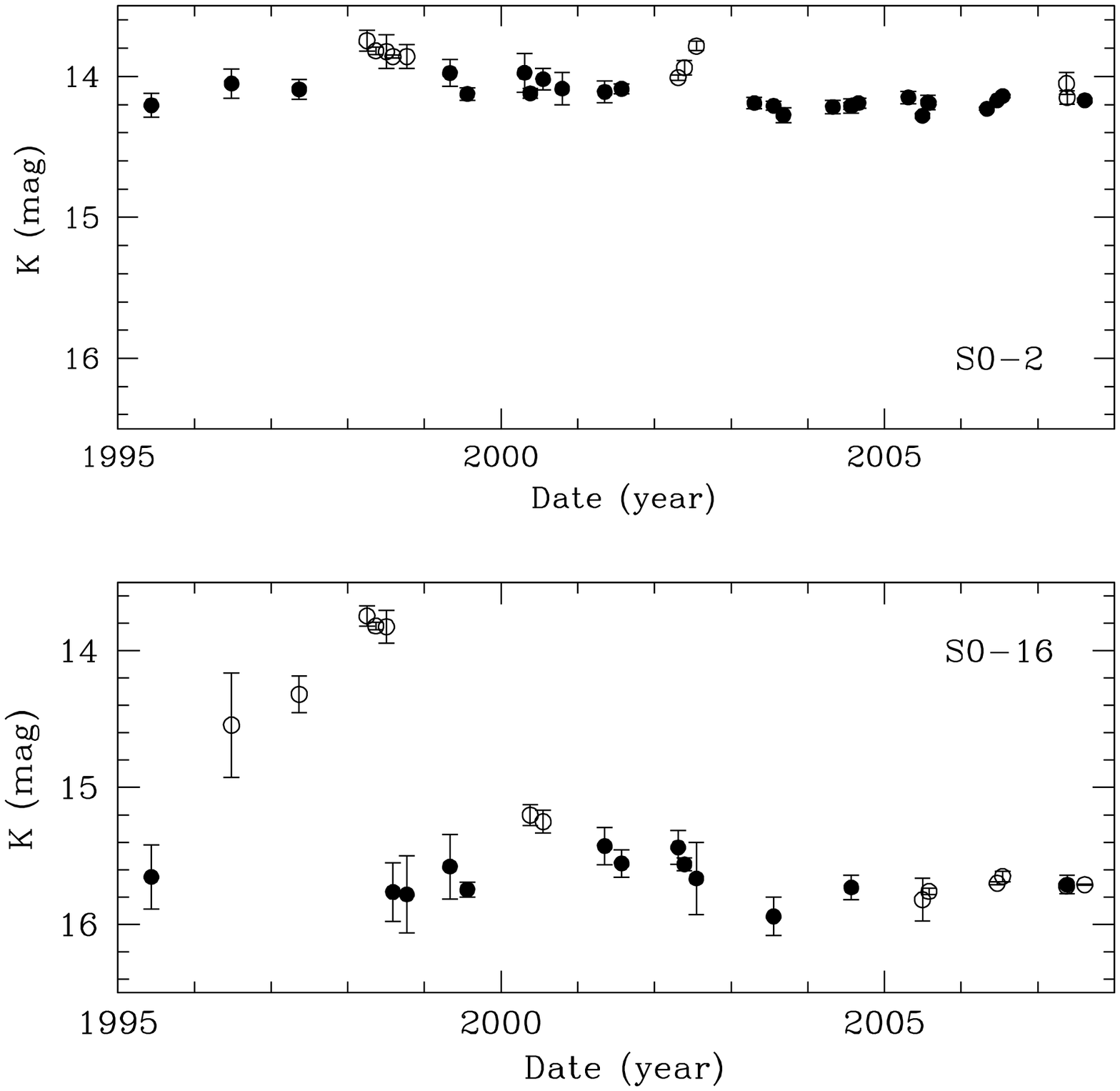}
\figcaption{
Photometric measurements vs. time for S0-2 (top) and S0-16 (bottom).  
Measurements that were made when these sources
coincided with another known source are plotted as unfilled points
and excluded from the model fitting procedure.  S0-16 is more affected
by underlying sources, because it is fainter.  
Even without {\it a priori} knowledge of the underlying sources, 
their effect is clearly visible in photometric measurements made in 1998 and 2002, for S0-2, and 1996-1999 \& 2000, for S0-16.
\label{s02_phot}
}
\end{figure*}

Source confusion can introduce positional biases that can be comparable to 
and, at certain times, larger than the statistical errors caused by
background or detector noise.  
This occurs when two stars are sufficiently close to 
each other that only one source, rather than two, is identified in our analysis
with a brightness that includes flux from both sources and a position that 
corresponds roughly to the photocenter of the two stars.  
We divide the problem of handling source confusion in our data set into
the following two cases: (1) the impact of unresolved, underlying stars that 
are  known sources, because they were sufficiently well separated at other 
times, and bright enough, to be independently detected, and (2) 
the impact of unresolved, underlying stars that are
not identified by this study at another 
time. Because the sources are moving so rapidly, instances of the former case 
are easily identified and are typically blended for one year. 
An underlying source that is comparably bright to the source of interest 
can have a significant impact on the astrometry;
to quantify this effect, we examine the idealized, noise-free case of a 
perfectly known PSF by using our empirical PSFs 
to generate idealized binary stars and running StarFinder on these simulated 
images, inputting the known PSF.  In this case, the astrometric bias is 
zero once the two components are detected.
As Figure \ref{bias_binary} shows, when the sources are blended, the resulting astrometric biases 
can be easily as large as 10 mas, which is much larger than our centroiding
uncertainties.  Such a large astrometric bias occurs when the 
underlying source is at least
half as bright as the primary source and has a projected, although
unresolved, separation of $\sim$ 40 mas.
We conservatively choose to eliminate all astrometric 
measurements that are known to be the blend of two sources 
from the orbital analysis; specifically, if the predicted positions
of two known sources are separated by less than 60 mas and only
one of them is detected, then that measurement is removed from our
analysis.
For S0-2 (K=14.2 mag), the eliminated data points are those made in 1998, due 
to confusion
with S0-19 (K=15.6 mag), in 2002, due to overlap with SgrA*-IR (K$_{median}$ = 16.4 mag, but can be as bright as 14 mag; see Do et al. 2008), 
and in
May 2007, due to superposition with S0-20 (K=15.9).  The impact of these overlapping
sources, in the first two cases, can be seen in the photometric measurements (see Figure \ref{s02_phot}).

Source confusion from unknown sources is a smaller effect than that from known
sources, since the unknown stars, in general, are fainter than the known 
sources.
Given the long time-baseline of the speckle imaging experiment,
knowledge of sources in this region is most likely complete
down to K= 16.0 mag.  While sources as faint as K = 19 mag
have been detected in this region with LGSAO,
crowding and the short time baseline of these deeper observations
limit the census of these sources.
Therefore, source confusion from unknown sources can
give rise to astrometric biases for S0-2 as large as 3 mas (from a K=16 mag 
source), but are typically 
significantly smaller since underlying sources will generally be
fainter than K = 16 mag.  To characterize the expected astrometric 
bias from the undetected source distribution, 
a Monte Carlo simulation was performed by generating multiple images with 
all known stars plus a random stellar 
distribution that, in total, follows the K luminosity function and radial 
profile from 
Sch\"odel et al. (2007).  By running these simulated images through 
our data analysis prodedure, we estimate that the astrometric error 
from unknown 
sources for S0-2 is, on average, 0.5 mas and 1.2 mas for the LGSAO and 
speckle images, respectively, and that it scales roughly with
the photometric bias and galacto-centric distance.  
However, it should be noted that the exact value of this bias is 
model dependent.  
While the photometric bias may be detected in the speckle data toward closest
approach (see Figure \ref{s02_phot}), the estimated astrometric biases 
are smaller than 
other sources of positional uncertainty already included for the majority of 
the S0-2 data points. 
We therefore do not incorpate them into the reported positional uncertainties.
Confusion with unknown sources gives rise to larger astrometric biases 
for S0-16, S0-19, and S0-20, since these sources are fainter than S0-2. 
Given the velocity dispersion in this region and the angular resolution
of the data sets, the expected timescale associated with biases from source 
confusion is $\sim$1-2 years.  

As a final step, the relative astrometric positions are placed in an absolute
coordinate reference frame using the positions of seven SiO masers
(Reid et al. 2003, 2007).  
Infrared observations of these masers with the Keck II LGSAO/NIRC2
system between 2005 and 2007 were obtained with the same
camera (i.e., plate scale) used for the precision astrometry measurements
described above, but with a nine position box pattern and a 6$\arcsec$ dither
offset to create a 22$\arcsec \times$22$\arcsec$ mosaic of these masers
(see Appendix C for details).
A comparison of the maser positions measured in this infrared
mosaic to the predicted radio positions at this epoch from
Reid et al. (2003) establishes that the mosaic has an average pixel scale of
9.963 $\pm$ 0.005 mas/pixel and a position angle of north with respect to the 
NIRC2 columns of 0.$^o$13 $\pm$ 0.$^o$02. 
This same analysis localizes the radio position of Sgr A* in the
infrared mosaic to within 5 mas in the east-west and 
north-south directions. 
By aligning the infrared
stars detected in both the larger infrared mosaic and the precision
astrometry image taken during the same observing run, we have
the necessary coordinate transformations to convert our 
relative astrometric position measurements into an absolute 
reference frame.  For the orbit analysis described in \S4,
the uncertainties in this transformation are applied
only after model orbits have been fit to the relative astrometry
and are a negligible source of uncertainty in the final
mass and R$_o$ estimates.

\subsection{Spectral Analysis \& Radial Velocities}

In the analysis of the spectral data, 
we accomplish the initial basic data processing steps
using standard IRAF procedures, for NIRC2 and NIRSPEC, and a facility 
IDL data extraction pipeline for OSIRIS.
Specifically, each data set is first (1) flat fielded, (2) dark
subtracted, (3) bad pixel and cosmic ray corrected, (4) spatially 
dewarped, and (5) wavelength calibrated.
Wavelength calibration is performed 
by identifying OH emission
lines from sky spectra and fitting a low-order polynomial function to
the location of the lines.
For the NIRSPEC spectra, neon emission lines from arc lamps provide
the wavelength calibration.  The accuracy of the wavelength
calibration is $\sim$9 km s$^{-1}$ or less for NIRC2 and OSIRIS as measured by 
the dispersion of the residuals to the fit.
Next, the one-dimensional stellar spectra are extracted using a spatial 
window that covers $\sim 0\farcs 1$ for the two dimensional spectral data sets
from NIRC2 and NIRSPEC.   For the three dimensional spectral data set from 
OSIRIS, an extraction box 0\farcs 14 $\times$ 0\farcs 14 was used.
To correct for atmospheric telluric absorption features,
each spectrum is divided by the spectrum of an A-type star.
Prior to this step, the A-type star's
strong intrinsic Br${\gamma}$ feature is removed.  In the case of the NIRC2 and OSIRIS
observations, this correction is done with observations of a G2V star, 
which is divided by a model solar spectrum.  The Br$\gamma$ corrected region in the G star is then substituted into the same region of the A star 
(Hanson et al. 1996).  In the 
case of the NIRSPEC observations, 
the A-type star's \brg feature is corrected with a model
spectrum of Vega\footnote{Model taken from the 1993 Kurucz Stellar Atmospheres 
Atlas 
(\url{ftp://ftp.stsci.edu/cdbs/cdbs2/grid/k93models/standards/vega\_c95.fits})}
rebinned to the resolution of the A-type star's spectrum and
convolved with a Gaussian to match the spectral resolution of the 
observations.  The resulting stellar spectra 
are corrected for all telluric absorption features; however, they
are still contaminated by background emission due
to the gas around the Galactic center.
The local background is estimated and removed by subtracting
spectra extracted from regions that are $\sim$0\farcs 1 away.
Finally, all the spectra within each night of observation
are combined in an average, weighted by the signal to noise ratio. 

\begin{figure*}
\epsscale{1.0}
\plotone{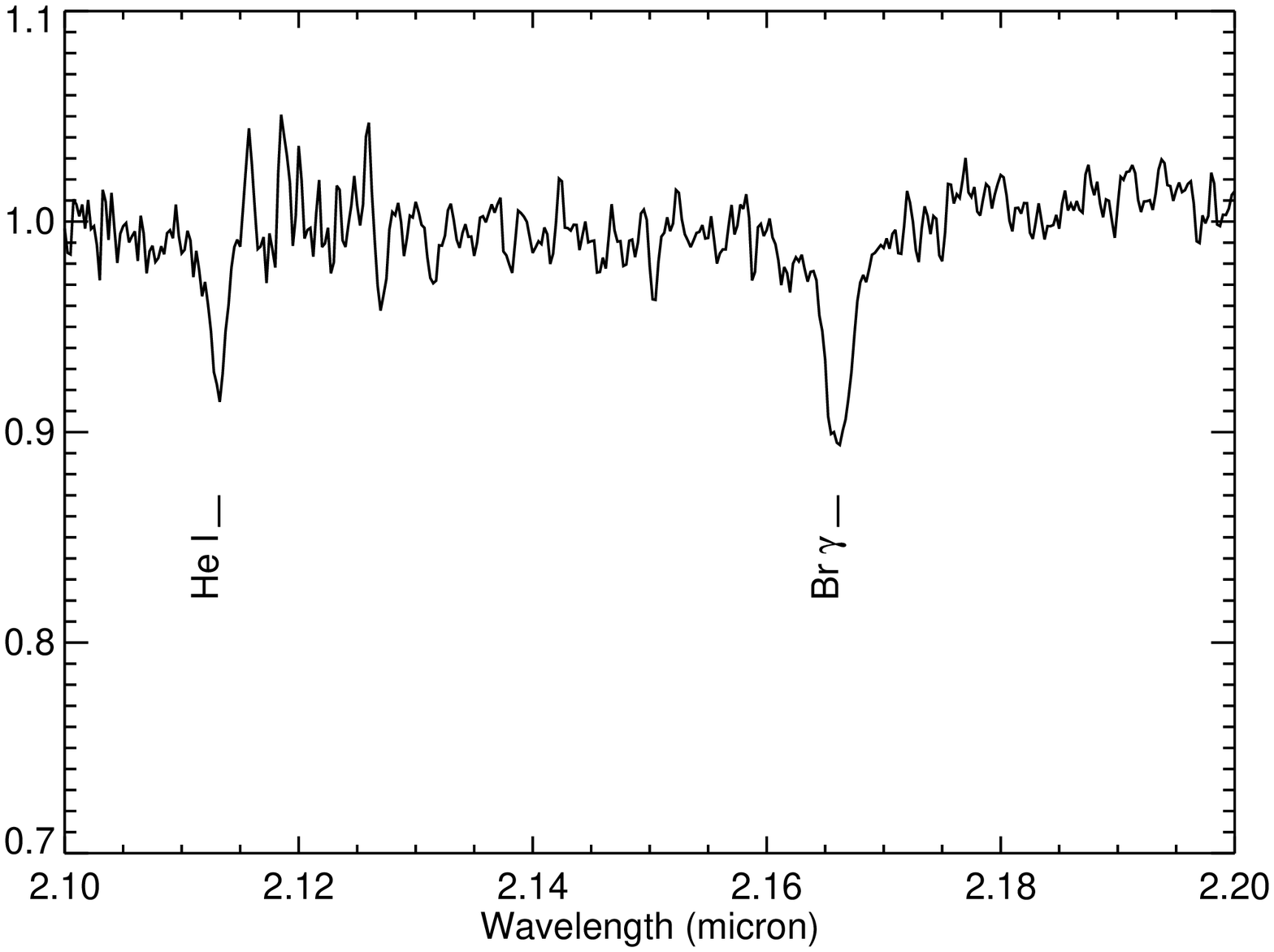}
\figcaption{The weighted average of all S0-2 spectra obtained with the W. M. Keck II
telescope.   Since only some of the data sets contain the shorter
wavelengths, the signal to noise ratio is lower at wavelengths shortward of 2.13 $\mu$m.
While \brg and He I lines are clearly detected, only the \brg line, 
which is stronger and not the blend of multiple lines, 
is used to measure the radial velocity of S0-2 as a function of time.
\label{spec_avg}
}
\end{figure*}

\begin{figure*}
\epsscale{0.8}
\plotone{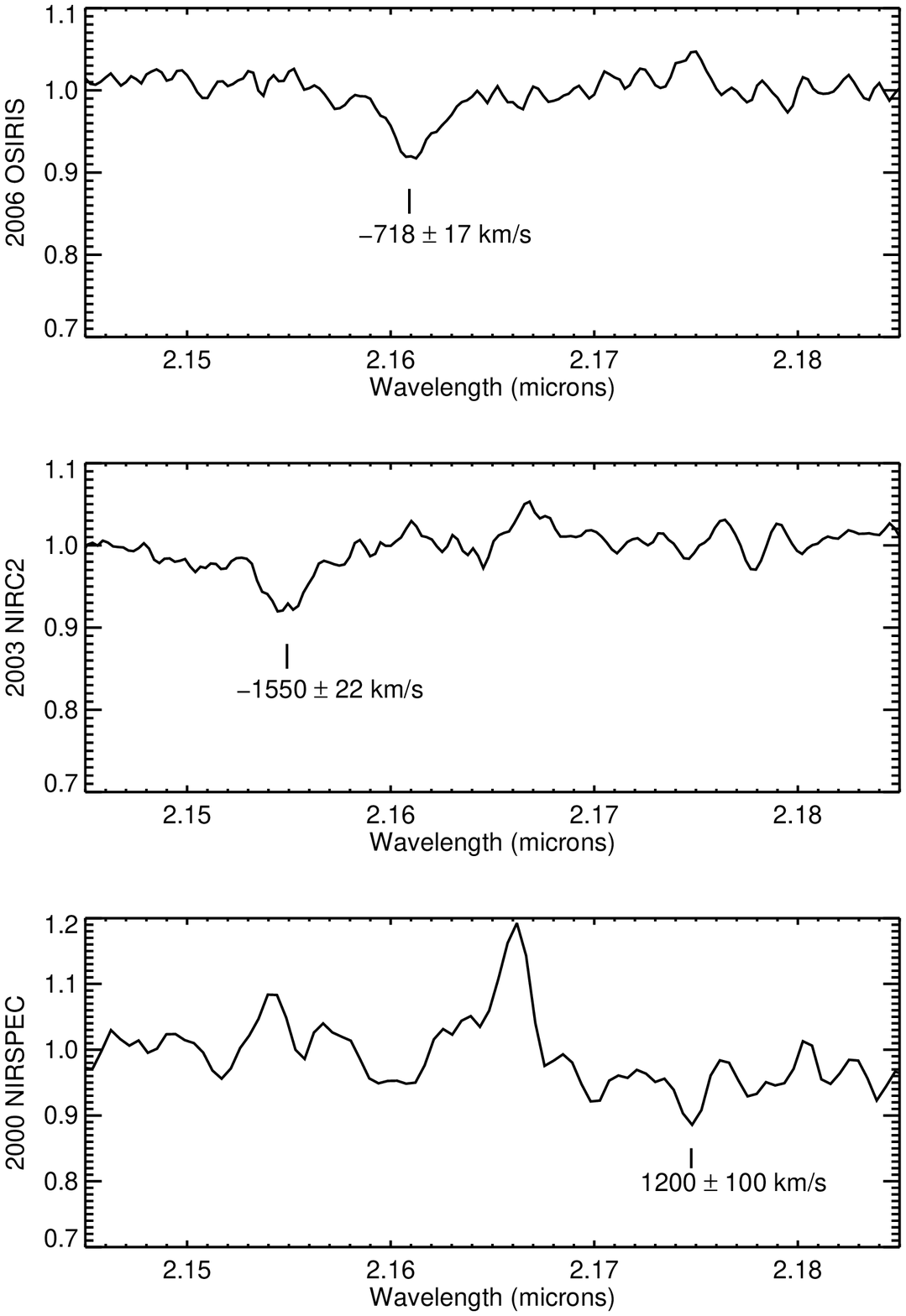}
\figcaption{
Measurements of S0-2's Br$\gamma$ line .  These three measurements 
show that, over time, S0-2's radial velocity has changed by more than
2,600 km s$^{-1}$.  
In order to improve the line detection in the 
low SNR NIRSPEC observation, 
the emission from the local gas was not removed, which leaves a large 
Br$\gamma$ emission feature centered at small radial velocities compared to 
that of the star at this time.  With improvements in the 
adaptive optics system and instrumentation 
(from NIRSPEC/NGSAO [bottom], to NIRC2/NGSAO [middle] and finally to 
OSIRIS/LGSAO [top]), 
the precision with which the Br$\gamma$ absoprtion line can be measured in S0-2 has improved by a factor of 5.
\label{spec_brg}
}
\end{figure*}

Radial velocity estimates are determined for each spectrum
on the basis of the location of the Br$\gamma$ line.
While a few of our spectra with
broader spectral coverage also show a weaker He I triplet at 2.116 $\micron$,
we do not incorporate measurements from this line, as it is a blend of 
transitions that can bias the resulting radial velocities (see Figure
\ref{spec_avg}).  A Gaussian model is fit to each of the 
Br$\gamma$ line profiles and the wavelength of the best fit peak, 
is compared to the rest wavelength of $\lambda_{vacuum}$ = 2.1661 $\micron$  to derive an observed 
radial velocity.  
To obtain radial velocities in the local standard
of rest (LSR) reference frame,
each observed radial velocity is corrected for the Earth's rotation, its
motion around the Sun, and the Sun's peculiar motion with respect to the LSR
(U = 10 km s$^{-1}$, radially inwards; Dehnen \& Binney 1998). 
Since the LSR is defined as the velocity of an object in circular orbit 
at the radius of the sun, the Sun's peculiar motion with respect 
to the average velocity of stars in its vicinity should give the Sun's motion 
toward the center of the Galaxy. 
The uncertainties in the final radial velocities are obtained from 
the rms of the fits to the line profile measurements from 
at least three independent subsets of the original data set.
Figure \ref{spec_brg} shows how S0-2's Br$\gamma$ line has shifted
over time and how the measurement of this line has improved by a factor
of 5 with improved instrumentation.
For the deep LGSAO spectroscopic observations, the radial velocity 
uncertainties for S0-2 are typically $\sim$20-25 km s$^{-1}$. 

\section{ORBITAL ANALYSIS \& RESULTS}
\label{sec:orbit}

\subsection{Point Mass Only Analysis}

To derive the black hole's properties, we assume that the stars are
responding to the gravitational potential of a point mass.
In this analysis, the 7 properties of the central black hole that are 
fitted are its mass ($M$), distance (R$_0$), location on the plane of the 
sky ($X_0$, $Y_0$) and motion ($V_x$, $V_y$, $V_z$).
In addition to these common free parameters, there are the following
6 additional free parameters for each star:
period ($P$), eccentricity ($e$),
time of periapse passage ($T_0$), 
inclination (i), 
position angle of the ascending node ($\Omega$), and
the longitude of periapse ($\omega$).
Using a conjugate gradient 
$\chi^2$ minimization routine that simultaneously fits 
the astrometric and radial velocity measurements,
we fit this model to measurements that are
given in Tables \ref{tbl_pos} \& \ref{tbl_rv},
which includes 27 epochs of astrometric measurements and 11 epochs of 
radial velocity (RV) measurements, as well as 5 additional epochs of 
radial velocity measurements reported in the literature
(Eisenhauer et al. 2003, 2005). This excludes all the astrometric 
measurements of S0-2 that are confused with another known 
source (see \S3.1).  While the 2002 astrometric data are eliminated due
to confusion with SgrA*, the 2002 RV points are not, since SgrA* is featureless
and therefore does not bias the measurement of RV from S0-2's Br$\gamma$
absorption line.
In total, there are 38 astrometric data points and 5 RV measurements.
All values reported for each parameter are the
best fit values obtained from minimizing the total $\chi^2$, 
which is the sum of the $\chi^2$ from each data type 
(i.e., $\chi^2_{tot} = \chi^2_{ast} + \chi^2_{RV}$).    

\begin{deluxetable*}{lrll}
\tabletypesize{\scriptsize}
\tablewidth{0pt}
\tablecaption{Summary of Keck Astrometric \& Photometric Measurements\label{tbl_pos}}
\tablehead{
	\colhead{UT Date} &
	\colhead{K$_{obs}$ (mag)} & 
	\colhead{X (mas)\tablenotemark{a}} &
	\colhead{Y (mas)\tablenotemark{a}} 
}
\startdata
1995.439 & 14.21 $\pm$ 0.09 & 	\phm{-}42.6 $~ \pm ~ $ 1.0 & 164.10 $\pm$ 0.98 \\
1996.485 & 14.05 $\pm$ 0.10 &	\phm{-}53.0 $~ \pm ~ $ 9.5 & 155.4 $ ~ \pm ~ $ 9.5 \\
1997.367 & 14.09 $\pm$ 0.07 &	\phm{-}56.5 $~ \pm ~ $ 1.7 & 137.0 $~ \pm ~ $ 1.7 \\
1999.333 & 13.98 $\pm$ 0.10 &	\phm{-}66.6 $~ \pm ~ $ 3.1 & \phm{0}91.5 $ ~ \pm ~ $ 3.1 \\
1999.559 & 14.12 $\pm$ 0.04 &	\phm{-}67.4 $~ \pm ~ $ 1.4 & \phm{0}88.3 $ ~ \pm ~ $ 1.4 \\
2000.305 & 13.98 $\pm$ 0.14 &	\phm{-}64.3 $~ \pm ~ $ 3.0 & \phm{0}65.8 $ ~ \pm ~ $ 3.1 \\
2000.381 & 14.12 $\pm$ 0.04 &	\phm{-}66.7 $~ \pm ~ $ 1.1 & \phm{0}63.0 $~ \pm ~ $ 1.1 \\
2000.548 & 14.02 $\pm$ 0.08 &	\phm{-}64.84 $ \pm   $ 0.78 & \phm{0}57.94 $\pm$ 0.80 \\
2000.797 & 14.09 $\pm$ 0.11 &	\phm{-}65.4 $~ \pm ~ $ 4.8 & \phm{0}46.8 $~ \pm ~ $ 4.9 \\
2001.351 & 14.11 $\pm$ 0.08 &	\phm{-}56.7 $~ \pm ~ $ 1.6 & \phm{0}26.5 $~ \pm ~ $ 1.6 \\
2001.572 & 14.09 $\pm$ 0.04 &   \phm{-}53.0 $~ \pm ~ $ 1.4 & \phm{0}14.2 $~ \pm ~$ 1.3 \\
2003.303 & 14.19 $\pm$ 0.04 &	-34.9 $~ \pm ~ $ 1.5 & \phm{0}69.5 $ ~ \pm ~ $ 1.6 \\
2003.554 & 14.21 $\pm$ 0.03 &	-35.45 $\pm$ 0.90 & \phm{0}81.04 $\pm$ 0.90 \\
2003.682 & 14.28 $\pm$ 0.05 &	-34.5 $~ \pm ~ $ 2.3 & \phm{0}87.4 $~ \pm ~ $ 2.3 \\
2004.327 & 14.22 $\pm$ 0.05 &	-32.15 $\pm$ 0.84 & 113.95 $\pm$ 0.86 \\
2004.564 & 14.21 $\pm$ 0.05 &	-28.7 $~ \pm ~$ 1.4 & 121.3 $~ \pm ~ $ 1.5 \\
2004.567 & 14.21 $\pm$ 0.02 &	-28.4 $~ \pm ~ $ 1.4 & 122.9 $~ \pm ~ $ 1.4 \\
2004.660 & 14.19 $\pm$ 0.04 &	-26.8 $~ \pm ~ $ 1.1 & 125.5 $~ \pm ~ $ 1.1 \\
2005.312 & 14.15 $\pm$ 0.04 &	-18.58 $\pm$ 0.88 & 142.43 $\pm$ 0.92 \\
2005.495 & 14.28 $\pm$ 0.02 &	-18.6 $~ \pm ~ $ 1.0 $\pm$ 1.1  & 145.3 $~ \pm ~ $ 1.0 $\pm$ 2.5  \\
2005.566 & 14.18 $\pm$ 0.05 &	-15.3 $~ \pm ~ $ 1.7 & 148.9 $~ \pm ~ $ 1.8 \\
2005.580 & 14.19 $\pm$ 0.01 &	-16.9 $~ \pm ~ $ 0.23 $\pm$ 1.0 & 146.8 $~ \pm ~ $ 0.23 $\pm$ 1.5 \\
2006.336 & 14.23 $\pm$ 0.01 &	\phm{0}-7.97 $\pm$ 0.13 $\pm$ 0.77 & 159.82 $\pm$ 0.13 $\pm$ 0.66  \\
2006.470 & 14.17 $\pm$ 0.01 &	\phm{0}-6.01 $\pm$ 0.14 $\pm$ 0.77 & 161.57 $\pm$ 0.14 $\pm$ 0.66  \\
2006.541 & 14.14 $\pm$ 0.01 &	\phm{0}-4.89 $\pm$ 0.17 $\pm$ 0.77 & 162.26 $\pm$ 0.17 $\pm$ 0.66  \\
2007.612 & 14.17 $\pm$ 0.01 &	\phm{-0}6.88 $\pm$ 0.20 $\pm$ 0.77  & 173.47 $\pm$ 0.20 $\pm$ 0.66  \\
\enddata
\tablenotetext{a}{X and Y are the relative positions in the EW and NS direction,
with increasing values to the E and N, respectively.  These values
are in our absolute coordinate system (i.e., relative to SgrA*-Radio; 
see Appendix C), but the uncertainties 
do not include the uncertainties in the absolute coordinate system.
Measurements that are confused with other known sources are not included
in this table. }
\tablenotetext{b}{Uncertainties from residual distortions in NIRC2 relative 
to the 2004 July reference image are reported separately (the second 
uncertainty term in the table) and should be added 
in quadrature to the other uncertainty terms to obtain the final positional
uncertainties; 
since the 2006 - 2007 LGSAO images are all obtained with the same set up, 
positions from these images have correlated residual distortion uncertainties.}
\end{deluxetable*}

\begin{deluxetable*}{llr}
\tabletypesize{\scriptsize}
\tablewidth{0pt}
\tablecaption{Summary of Keck Radial Velocity Measurements\label{tbl_rv}}
\tablehead{
	\colhead{UT Date} &
	\multicolumn{2}{c}{Radial Velocity (km s$^{-1}$)} \\
	\colhead{} &
	\colhead{Observed} &
	\colhead{LSR} 
}
\startdata
2000.487  &  1192 $\pm$ 100  &	1199  $\pm$ 100   \\
2003.438  & -1556 $\pm$ 22   &	-1550 $\pm$  22   \\
2004.474  & -1151 $\pm$ 57   &	-1143 $\pm$  57   \\
2005.410  &  -945 $\pm$ 16   &	-926  $\pm$  16   \\
2005.504  &  -853 $\pm$ 31   &	-850  $\pm$  31   \\
2006.391  &  -715 $\pm$ 21   &	-692  $\pm$  21   \\
2006.461  &  -728 $\pm$ 17   &	-718  $\pm$  17   \\
2006.495  &  -699 $\pm$ 36   &	-695  $\pm$  36   \\
2006.497  &  -717 $\pm$ 37   &	-713  $\pm$  26   \\
2007.385  &  -507 $\pm$ 50   &	-483  $\pm$  50   \\
2007.548  &  -502 $\pm$ 50   &	-506  $\pm$  50   \\
\enddata
\end{deluxetable*}

The uncertainties on the fitted parameters are estimated using a Monte Carlo 
simulation, which is a robust approach when performing a fit with many correlated 
parameters.  
We created $10^5$ artificial datasets ($N_{sim}$) containing as many points as the observed 
dataset (astrometry and radial velocities), 
in which each point is randomly drawn from a Gaussian distribution centered on 
the actual measurement and whose 1 $\sigma$ width is given by the associated 
uncertainty, 
and run the $\chi^2$ minimization routine for each realization.  
$N_{sim}$ was set to $10^5$ in order to achieve $\sim$6\% 
accuracy in the resulting estimates of the 
a 99.73\% confidence limits (3$\sigma$ equivalent for a gaussian distribution) 
of the orbital parameters.
Because the $\chi^2$ function contains many local minima, each realization
of the data is fit 1000 times ($N_{seed}$) with different seeds to find the 
global minimum.   
The resulting distribution of $10^5$ values of the fitted parameters
from the Monte Carlo simulation, once normalized, is the
a joint probability distribution function of the orbital parameters 
($PDF({\vec O})$, where $\vec O$
is a vector containing all the orbital parameters,$O_i$).
For each orbital parameter, $PDF({\vec O})$ is marginalized against all
other orbital parameters to generate a $PDF(O_i)$.  The confidence limits
for 
each parameter are obtained by integrating
each $PDF(O_i)$ from its peak\footnote{While the best values
from minimizing $\chi^2$ can differ slightly (but well within
the uncertainties) from the peak of the $PDF(O_i)$ values, this has negligible
impact on the reported uncertainties.}
outwards to a probability of 68

\begin{figure*}
\epsscale{1.0}
\plotone{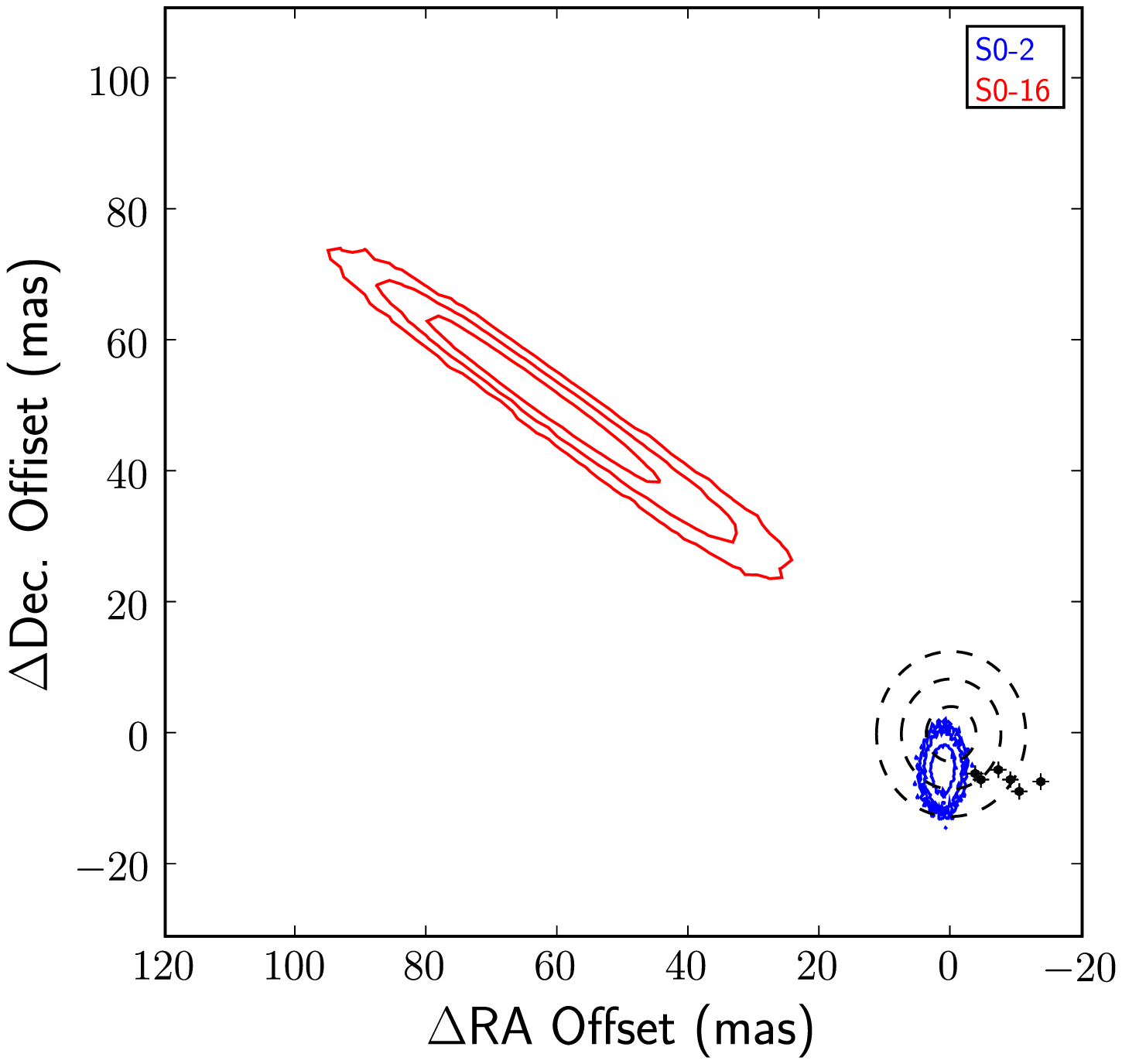}
\figcaption{
Comparison of estimates of the black hole's location.
Colored contours represent the estimates 
of the dynamical center from model fits to kinematic measurements of
S0-2 (K = 14.0; blue) and
S0-16 (K = 15.0; red).  Black contours show the SgrA*-Radio position.
All contours are plotted at the 68\%, 95\%, and 99.7\% confidence levels
(equivalent to 1, 2, and 3$\sigma$ for a Gaussian distribution).
The solid black points are all the measurements of SgrA*-IR (K$\sim$16) 
in the maps used
for the astrometric analysis.  
The discrepancy in the black hole's location from S0-16's positional 
measurements appear to be a consequence of biases from unrecognized,
underlying stars and thus only S0-2's measurements are used to 
infer the properties of the central black hole.  
Likewise, the astrometric positions of SgrA*-IR, which is even
fainter than S0-16, also may be biased (see discussion in \S4.1) and 
are therefore not used to constrain the orbital model used
to fit S0-2.
\label{bhPosition}
}
\end{figure*}

Compared to all other stars at the center of the Milky Way, 
S0-2 dominates our knowledge of the central black hole's properties.
Two facts contribute to this effect.  Most importantly, it has
the shortest known orbital period 
(P = 15 yr; Sch\"odel et al. 2002, 2003; Ghez et al. 2003, 2005a).
Furthermore, among the known short-period stars, it is the brightest star 
and therefore the least affected by stellar confusion (see Figure 1).  
Several other stars, in principle, also offer constraints on 
the black hole's properties.  In particular, 
S0-16 is the next most kinematically important star, as it is the only
other star that yields an independent solution for the black hole's properties.
However, independent solutions for the black hole's position from
fits to S0-2 and S0-16 measurements differ by more than 5 $\sigma$
(see Figure \ref{bhPosition}).
While S0-16's measurements in 2000 have already been omitted due
to overlap with the position of SgrA*, three independent lines of 
reasoning lead us to believe that some of S0-16's remaining 
astrometric measurements must be significantly biased by radiation from 
unrecognized, underlying stars.  
First, as shown in Figure \ref{bias_binary}, unknown sources
can introduce astrometric biases as large as 9 mas for S0-16 (K=15),
in contrast with only 3 mas for S0-2 (K = 14),
because it is only 1 mag
above the completeness limit for detection in the speckle data set
(K$\sim$16 mag; see \S3.1). 
Second, a comparison of the solution for the position of the 
black hole ($X_0, Y_0$) based on both the astrometric and radial velocity 
measurements
to that based on astrometry alone (fixing the distance, which cannot
be solved for without radial velocities) yields a consistent position
from modeling the two cases for S0-2, but produces different results for
the two cases from modeling S0-16's measurements, with the inferred
$X_0$ and $Y_0$ from astrometry alone shifting further away from that 
obtained from modeling S0-2's orbit prediction and thereby increasing the discrepancy 
to 10$\sigma$.
Third and last, while the position of the dynamical center 
from S0-2's orbit is statistically
consistent with SgrA*-Radio/IR, which is the emissive source 
associated with the central black hole (e.g., Melia \& Heino 2001; Genzel et al. 2003a; Ghez 
et al. 2004; 2005b; Hornstein et al. 2007), the solution from S0-16 is not
(see Figure \ref{bhPosition});
this 
difference cannot be explained by allowing the black hole to move with time
or by introducing an extended mass distribution.
We therefore restrict our remaining analysis to S0-2.

\begin{figure*}
\epsscale{0.8}
\plotone{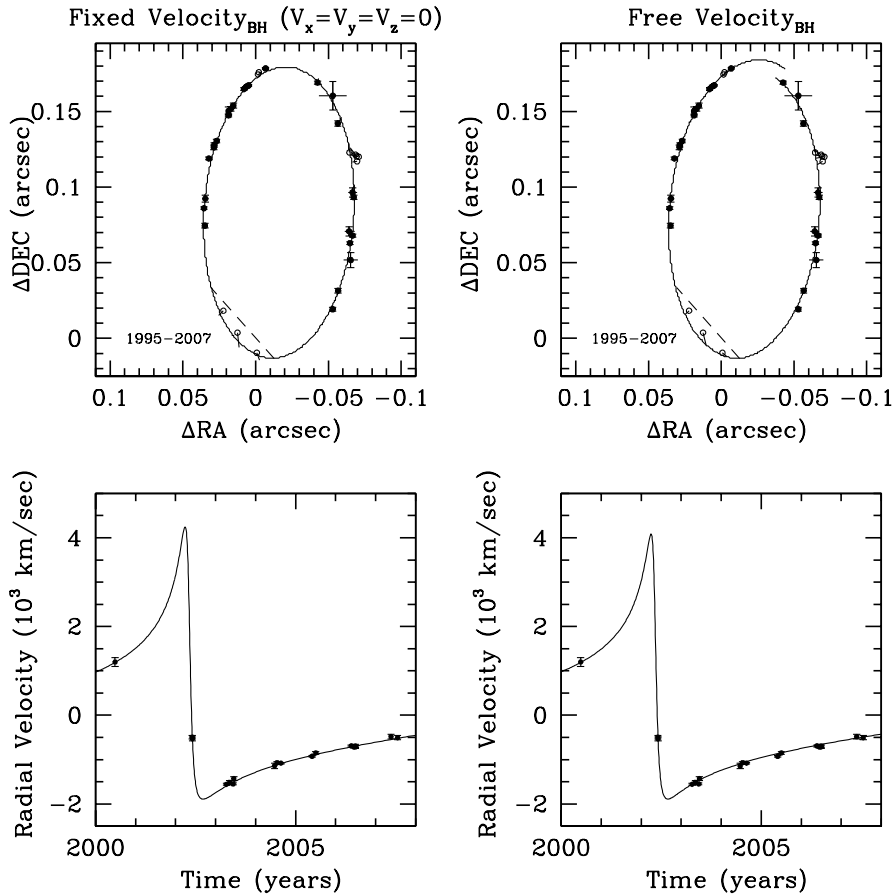}
\figcaption{
The best fit to the astrometric and radial velocity data, assuming a
Keplerian orbital model.  
The filled points were included in the formal fit, while the unfilled 
points are measurements that are excluded due to source confusion.  
Uncertainties are plotted on all points, except the unfilled/excluded 
points (here the uncertainties are comparable to the size of the points)
for clarity.  
({\it Left}) 
To compare with what has been done in the past to estimate R$_0$, we show
the fit to the data with a 10 parameter model, which 
includes the black hole's mass (M$_{bh}$), distance (R$_0$), and 
location in the plane of the sky ($X_0, Y_0$)
as free parameters, and which fixes the black hole's three dimensional 
velocity ($V_x, V_y, V_z$) to zero.  This 
results in a $\chi^2/dof \sim 1.4$. 
({\it Right}) 
The data are better reproduced by a 13 parameter model, which 
includes the black hole's mass (M$_{bh}$), distance (R$_0$), location in the plane of the sky ($X_0, Y_0$), and three dimensional velocity ($V_x, V_y, V_z$) 
as free parameters, and results in a $\chi^2/dof \sim 0.97$. 
Adding these extra free parameters, and in particular $V_z$, increases
the uncertainties in the black hole's properties by a factor of two. 
\label{s02_orb}
}
\end{figure*}

\begin{figure*}
\epsscale{0.8}
\plotone{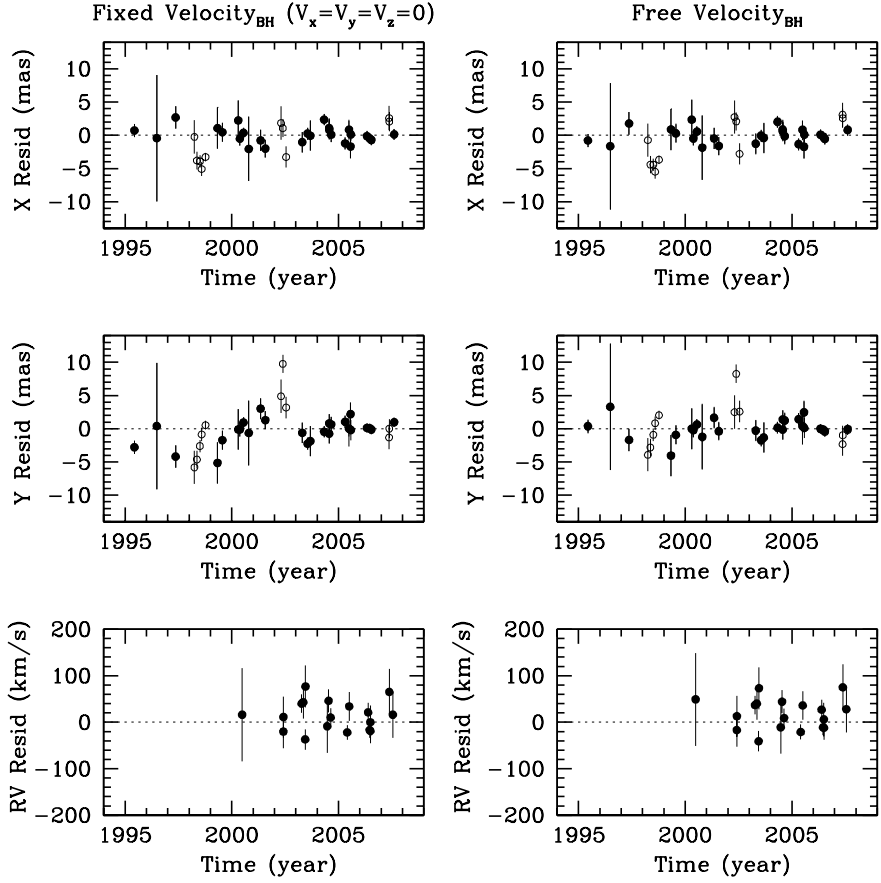}
\figcaption{
The residuals to the best fit Keplerian orbital models shown in Figure 9.  
The filled points were included in the formal fit, while the unfilled 
points are measurements that are excluded due to source confusion.  
\label{s02_orb_resid}
}
\end{figure*}

\begin{figure*}
\epsscale{1.0}
\plotone{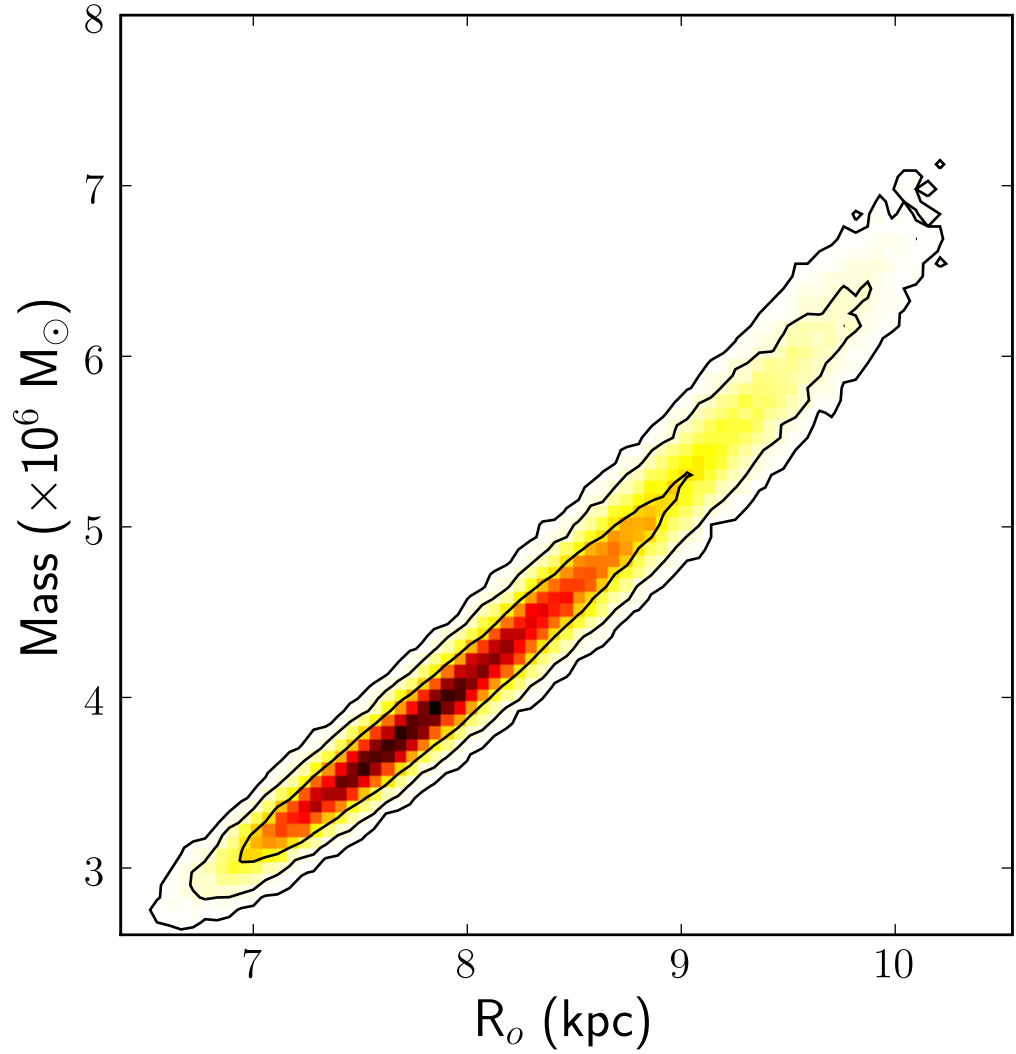}
\figcaption{
The correlation of the estimated black hole's mass and distance.
The density of solutions from the MC simulations are shown as a color 
image, with the contours marking the 68\%, 95\%, and 99.7\% confidence limits.
While mass and distance are well determined from the orbit of S0-2, 
they are not independent quantities.  The exact scaling depends on
the relative impact of the astrometric and radial data on the model
fits.  Currently, the inferred  mass scales with the inferred
distance as M $\propto$ R$_0$$^{1.8}$.
\label{massRo}
}
\end{figure*}

As shown in Figures \ref{s02_orb} \& \ref{s02_orb_resid}, the astrometric and 
radial velocity measurements for S0-2 are well fit by a simple 
Keplerian model.  For a 13 parameter model (right-hand side of figures), the best fit to the data 
produces a total $\chi^2$ of 54.8 for 57 degrees of freedom ($dof$)
and a $\chi^2/dof$ of 0.961.
From the Monte Carlo simulation, we derive  
probability distributions for the central black hole's
properties, which are shown in Figure \ref{massRo} and characterized
%
%
in Table \ref{tbl_orb}.  These distributions give a best fit
for the central black hole's mass of M$_{bh}$ = $4.1 \pm 0.6 \times 10^6  M_{\odot}$ and
distance of R$_0$ = 8.0 $\pm$ 0.6 kpc (all quoted uncertainties are 68\% confidence values).  The position of the black hole is 
confined to within $\pm$ 1 mas ($\sim$100 Schwarzschild Radii).  
As can be seen in Figure \ref{massRo}, 
the inferred black hole's mass is highly correlated with its distance.
Estimates from orbital modeling are expected to have a power law relationship 
of the form $Mass \propto M_{\odot} ~ Distance^{\alpha}$ with $\alpha$ between 1 and 3.
For the case of astrometric data only, $\alpha$ should be 3 and, for
the case of radial velocity data only, $\alpha$ is expected to be 1.  
Currently, the relationship is $M = (4.1 \pm 0.1 \times 10^6 M_{\odot})(R_0/8.0 ~ kpc)^{1.8}$, which suggests that 
the astrometric and radial velocity data sets are having roughly 
equal affect in the model fits for mass\footnote{The uncertainty in the
mass scaling relationship is obtained for the case in which R$_0$ is
fixed to 8.0 kpc and therefore does not include the uncertainty in R$_0$.}.  

\begin{deluxetable*}{lll}
\tabletypesize{\footnotesize}
\tablewidth{0pt}
\tablecaption{Orbital Elements for S0-2 and the Implied Black Hole Properties\label{tbl_orb}}
\tablehead{
	\colhead{Parameter (Symbol) [Unit]} &
	\colhead{V$_z = $0 Case\tablenotemark{a}} & 
	\colhead{V$_z$ Unconstrained Case} \\
}
\startdata
Distance (R$_o$) [kpc]\tablenotemark{b}				& \phm{-000}8.36\phm{00} $\pm$ $_{0.44}^{0.30}$ & \phm{-000}7.96\phm{00} $\pm$ $_{0.70}^{0.57}$ \\
Period (P) [yrs] 				& \phm{-00}15.78\phm{00} $\pm$ 0.35 	& \phm{-00}15.86\phm{00} $\pm$ $_{0.45}^{0.10}$  \\
Semi-major axis (a) [mas]			& \phm{-0}124.4\phm{000} $\pm$ $_{3.3}^{2.4}$	& \phm{-0}126.5\phm{000} $\pm$ $_{5.0}^{1.8}$  \\
Eccentricity (e)				& \phm{-000}0.8866 $\pm$ 0.0059 	& \phm{-000}0.8904  $\pm$ $_{0.0075}^{0.0051}$ \\
Time of Closest Approach (T$_o$) [yr] 		& \phm{-}2002.3358 $\pm$ $_{0.0093}^{0.0065}$  & \phm{-}2002.342\phm{0} $\pm$ 0.010  \\
Inclination (i) [degrees]			& \phm{-0}135.3\phm{000} $\pm$ 1.3 & \phm{-0}134.6\phm{000} $\pm$ 1.3 \\
Position Angle of the Ascending Node ($\Omega$) [degrees] & \phm{-0}225.9\phm{000} $\pm$ 1.3 & \phm{-0}226.44\phm{00} $\pm$ $_{1.4}^{0.71}$ \\
Angle to Periapse ($\omega$) [degrees]			& \phm{-00}65.18\phm{00} $\pm$ 1.2 & \phm{-00}66.0\phm{000} $\pm$ $_{1.7}^{1.1}$ \\
X Dynamical Center (X$_0$ - X$_{SgrA*-Radio}$) [mas]\tablenotemark{b,c} & \phm{-000}0.95\phm{00} $\pm$ $_{1.4}^{0.46}$ & \phm{-000}1.49\phm{00} $\pm$ $_{0.87}^{1.1}$ \\
Y Dynamical Center (Y$_0$ - Y$_{SgrA*-Radio}$) [mas]\tablenotemark{b,c} &  \phm{000}-4.8\phm{000} $\pm$ $_{1.6}^{2.2}$ & \phm{-00}-5.4\phm{000} $\pm$ 2.0 \\
X Velocity ($V_x$) [mas/yr]				& \phm{000}-0.40\phm{00} $\pm$ 0.25 & \phm{-00}-0.47\phm{00} $\pm$ $_{0.33}^{0.12}$\\
Y Velocity ($V_y$) [mas/yr]				& \phm{-000}0.39\phm{00} $\pm$ $_{0.18}^{0.09}$ & \phm{-000}0.36\phm{00} $\pm$ 0.12 \\
Z Velocity ($V_z$) [km/sec]				& \nodata & \phm{00}-20\phd \phm{0000} $\pm$ $_{37}^{29}$ \\
\hline
Mass (M$_{bh}$) [$10^6$ M$_{\odot}$] 		& \phm{-000}4.53\phm{00} $\pm$ $_{0.55}^{0.34}$ & \phm{-000}4.07\phm{00} $\pm$ $_{0.78}^{0.52}$  \\
Density ($\rho$) [$10^{15}$ M$_{\odot}$pc$^{-3}$] & \phm{-000}5.83\phm{00} $\pm$ $_{0.97}^{0.28}$  & \phm{-000}6.3\phm{000} $\pm$ $_{1.4}^{0.56}$  \\	
Periapse Distance (R$_{min}$) [mpc]		& \phm{-000}0.570\phm{0} $\pm$ 0.037 & \phm{-000}0.535\phm{0} $\pm$ $_{0.071}^{0.049}$ \\
\enddata
\tablecomments{Parameters below
the horizontal line are derived from those above the line and are provided for convenience.}
\tablenotetext{a}{
Allowing for the uncertainty in the LSR in $V_z$ ($\pm$ 2 km/sec; Gould 2004) produces 
results that are not distinguishable from those reported above for the $V_z$ = 0 case. }
\tablenotetext{b}{The reference time for the position of the black hole, when the velocity is a free parameter, is 2000.0}  
\tablenotetext{c}{Uncertainties in the position of SgrA*-Radio are not incorporated into the 
uncertainties of $X_0$ and $Y_0$. } 
\end{deluxetable*}

A fit that includes the biased astrometric data points significantly
alters the best fit solution for S0-2.  
Including both the 1998 and 2002 data points,
which correspond to confusion with S0-19 and SgrA*-IR respectively,
results in a higher mass ($5.7 \times 10^6 M_{\odot}$), distance
(9.4 kpc), and $\chi^2/dof$ (1.7).  
Including the 2002 but not the 1998 data points 
also produces elevated values ($5.2 \times 10^6 M_{\odot}$ and 9.1 kpc)
and $\chi^2/dof$ (1.1). 
%
This demonstrates that it is important to account for the astrometric
biases introduced by unresolved sources.

Formal uncertainties in mass and distance estimates from orbital
fits can be reduced by adding {\it a priori} information.  In particular,
it is, in principle, possible to constrain the dynamical center to be at 
the position of SgrA*-IR.  
However, as shown in Figure \ref{bhPosition}, the six measurements of 
SgrA*-IR's position in the 
deep LGSAO images (2005-2007), which have the most precise 
astrometric measurements, have an average value that differs from the position
of the black hole inferred from S0-2's orbit by 9.3 mas and 
a variance of 3 mas, which is a factor of 4 larger than expected from
the measured positional uncertainties (0.7 mas).
SgrA*-IR is located where the underlying sources are expected to 
have the highest number 
density and velocity dispersion, which should induce time variable
positional biases.   SgrA*-IR's average K magnitude in these deep LGSAO 
images is 16.4,
which is comparable to the completeness limit for sources in this
region (see \S3.1) and which is, consequently, potentially subject to 
large astrometric biases (see Figure \ref{bias_binary}).
We therefore suspect that the measured positions of SgrA*-IR
suffer from astrometric biases from underlying sources and do not use
its positions to constrain the model fits.  

Another prior, which has been imposed in earlier orbital analyses of
S0-2 for R$_0$ (Eisenhauer et al. 2003; 2005), is on the black hole's motion relative to 
the measurements' reference 
frame.  
Setting the three 
dimensional velocity to zero and fitting a 10 parameter model 
($\chi^2/dof$ = 1.3; see left-hand side of Figures \ref{s02_orb} \& \ref{s02_orb_resid}) yeilds uncertainties in the 
black hole's properties that are a factor of 
2 smaller (R$_0$ = 8.0 $\pm$ 0.3 kpc and M$_{bh}$ = $4.4 \pm 0.3 \times 10^6 M_{\odot}$).
However this assumption is not justified 
(see, e.g., Salim \& Gould 1999; Nikiforov 2008).  
Introducing $V_x$ and $V_y$ (defined such that positive numbers are motions
in the E and N directions, respectively) into the fit allows the
dynamical center to move linearly in time in the plane of the sky with
respect to the cluster reference frame.
Such an apparent motion can arise from either
a physical or a data analysis effect.  In the case of a physical effect,
the black hole could be moving with respect to the stellar cluster
under the gravitational influence of a massive companion or the black hole
and the cluster could be participating in a mutually opposing sloshing mode.
In the case of a data analysis effect, the reference frame could be 
non-stationary with respect to the position of the dynamical center, which 
might arise if there was a systematic problem in our alignment
of images.  Introducing these two parameters therefore provides a way 
of examining possible systematic reference frame problems.
Fits to a 12 parameter model ($V_z$ fixed to zero) to the data have a minimum
$\chi^2/dof$ of 0.95, uncertainties in the
black hole's properties that are larger than the 10-parameter model, but
smaller than the 13-parameter model
(R$_0$ 8.4 $\pm$ 0.4 kpc and M$_{bh}$ = $4.5 \pm 0.4 \times 10^6 M_{\odot}$),
and an estimate for the black hole's motion relative to the central 
stellar cluster of $V_x$ = -0.40 $\pm$ 0.25 mas/yr (17 $\pm$ 11 km/sec) 
and $V_y$ = 0.39 $\pm$ 0.14 mas/yr (16 $\pm$ 6 km/sec).  
Since these relative velocities are comparable to the constraints on the 
IR reference frame's motion with respect to SgrA*-Radio (i.e., an absolute 
reference frame in which the black hole's position is known; see Appendix C), 
it is important to leave $V_x$ and $V_y$ as free parameters, even for
the case in which one assumes that the black hole has no intrinsic motion motion
with respect to the cluster.  Because the black hole is so often assumed 
to be at rest, we report the complete solution for the 12 parameter fit
(V$_z$ fixed to zero) in Table \ref{tbl_orb}.

\begin{figure*}
\epsscale{1.0}
\plotone{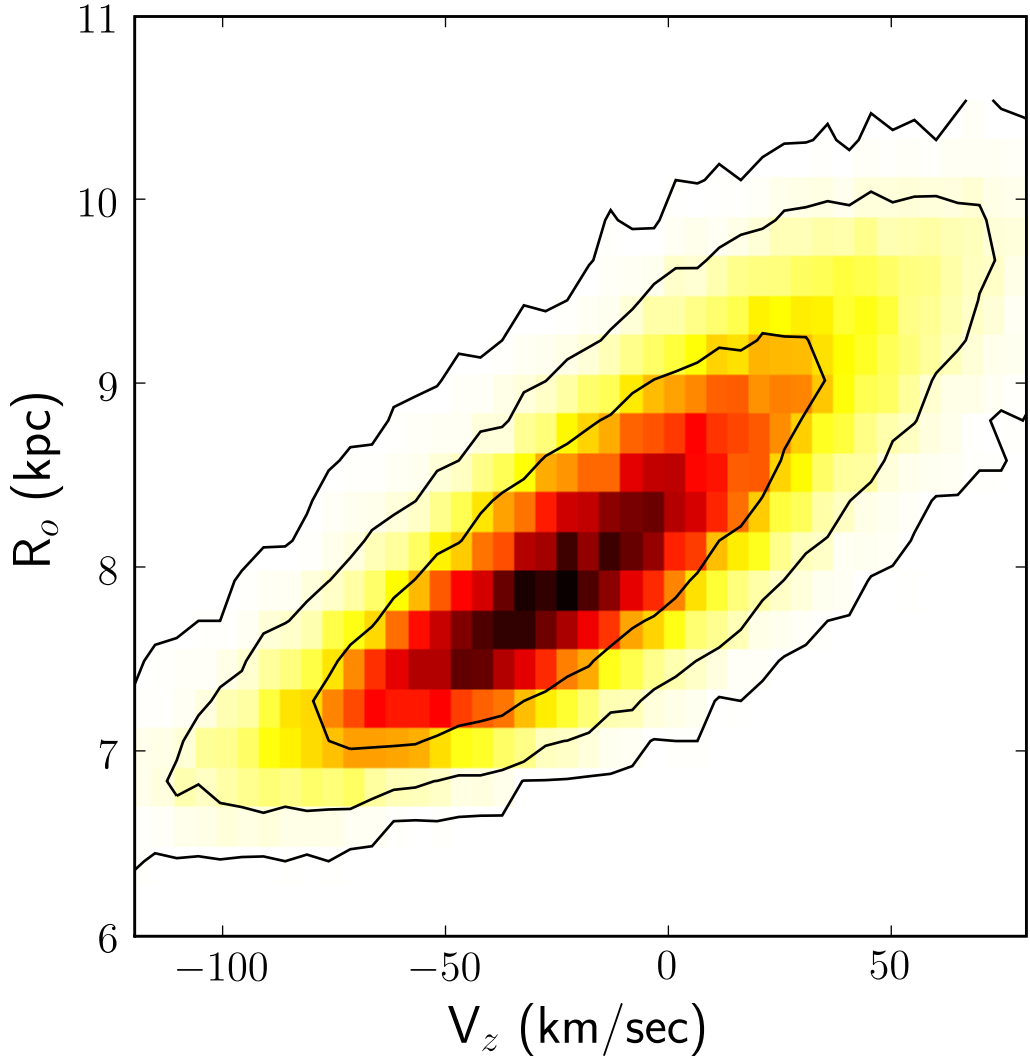}
\figcaption{
Correlation of the estimated black hole's distance and line-of-sight
velocity ($V_z$) from our 13 parameter model fit. 
$V_z$ dominates the uncertainties in R$_0$ and consequently M$_{bh}$.
Priors on $V_z$ can reduce the uncertainties in R$_0$ by a factor
of two.  All previous studies have set $V_z$ to zero, which 
implicitly assumes that there are no massive companions to our Galaxy's central 
supermassive black hole and that the local standard of rest is perfectly known.  
\label{RoVz}
}
\end{figure*}

As Figure \ref{RoVz} shows, the black hole's motion along the line of sight
with respect to our assumed local standard of rest ($V_z$) 
dominates the uncertainties in R$_0$ in our 13 parameter model.  Priors on $V_z$ therefore
have a signficant impact on the resulting uncertainties.  Unlike the plane of the sky,
the reference frame along the line of sight is unlikely to have an
instrumental systematic drift, since each of the spectra are calibrated against 
OH lines (see \S3.2).  
However, it is 
possible that there is a residual radial velocity offset between the LSR and
the S0-2 dynamical center.  The Sun's peculiar motion
with respect to the LSR along the line of sight might differ from the assumed
10 km s$^{-1}$; that is, the practical realization of the LSR is not on
a circular orbit around the Galactic center as might occur due the bar 
potential or to the spiral perturbations, so that the average velocity of 
stars in the solar vicinity might have a (small) net radial component.
Alternatively, the dynamical 
center of S0-2 could differ from the dynamical center of the Galaxy as determined
at the Sun's (i.e., LSR's) distance, as might result from the presence of an
intermediate mass black hole companion. 
From the model fit, the implied motion of the LSR along the line-of-sight
with respect to S0-2's dynamical center is 
-20 $\pm$ 33 km/sec, which is consistent with
no net motion.  While no significant motion is detected in 
$V_x$, $V_y$, or $V_z$, the 3$\sigma$ upper limits for the magnitudes of 
all three are comparable to one another in our 13 parameter model 
(48, 30, and 119 km/sec, respectively).
Since there are no direct contraints on these quantities that can improve 
these limits, we have allowed them to be fully free parameters.  However, 
if we {\it assume} that the black hole is stationary with respect to the
Galaxy, we also need to consider the case of Vz set to zero\footnote{Allowing 
for the 
uncertainty in the LSR in $V_z$ ($\pm$ 2 km/sec; Gould 2004) produces
results that are not distinguishable from those reported for the $V_z$ = 0
case.}.

\subsection{Point Mass Plus Extended Mass Distribution Analysis}

\begin{figure*}
\epsscale{1.0}
\plotone{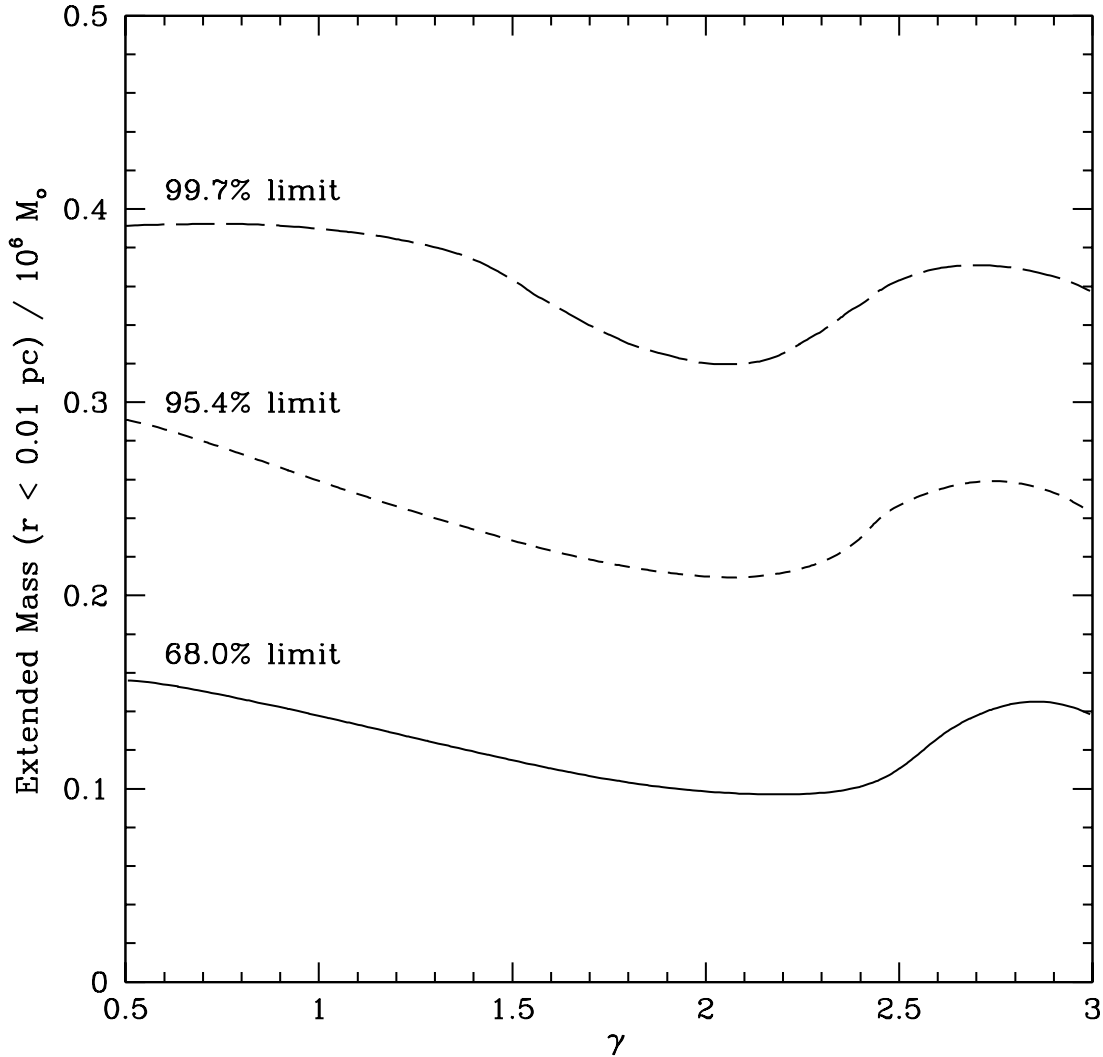}
\figcaption{
Limits on the amount of mass in an extended distribution contained within 
S0-2's apoapse distance.  The three lines correspond to the 68.3\%, 95.4\%, and
99.7\% upper-bound confidence limits.  
The 99.7\% confidence upper-bound of 3-4 $\times$ 10$^5$ M$_{\odot}$ 
is a fairly weak function of the slope of the assumed power-law mass profile.
Simple models of the stellar distribution suggest 
M$_{ext}(< 0.01 \textrm{pc}) \sim 10^3 M_{\odot}$, a factor of $\sim$100
smaller than the current measurement uncertainty.
\label{extended}
}
\end{figure*}

Limits on an extended mass distribution within S0-2's orbit
are derived by assuming that the gravitational potential consists of a point 
mass and an extended mass distribution, and allowing for a 
Newtonian precession of the orbits (see, e.g., Rubilar \& Eckart 2001).
In order to do this, we use the orbit fitting procedure 
described in Weinberg et al. (2005), and adopt an extended mass distribution
that has a power-law density profile $\rho(r)=\rho_0(r/r_0)^{-\gamma}$.  This
introduces two additional parameters to the model: the normalization of the 
profile and its slope $\gamma$. The total enclosed mass is then given by
\begin{equation}
M(<r) = M_{\rm BH} + M_{\rm ext}(<r_0) \left(\frac{r}{r_0}\right)^{3-\gamma},
\end{equation}
where we quote values for the normalization 
$M_{\rm ext}(<r_0)$ at $r_0=0.01\textrm{ pc}$, corresponding to the 
characteristic scale of the orbit.  Figure \ref{extended} shows 
the constraint 
on $M_{\rm ext}(<0.01\textrm{ pc})$ and $\gamma$ from a fit to the astrometric
and radial velocity measurements for S0-2.
The 99.7\% confidence upper-bound on the extended mass is $M_{\rm ext}(<0.01\textrm{ pc})\simeq 3-4 \times 10^5 M_\odot$ and has only a weak dependence on $\gamma$.

Mouawad et al. (2005) report a similar upper-bound on the extended mass in fits 
to the orbit of S0-2.  Their analysis differs only slightly from that 
presented here in that it forces the focus to be at the inferred radio position of 
Sgr A*, assumes a Plummer model mass distribution, and 
is based on data presented in Eisenhauer et al. (2003).
Similarly, 
Zakharov et al. (2007) use an order of magnitude analysis to show that if the 
total mass of the extended matter enclosed within the S0-2 orbit is 
$\ga 10^5 M_\odot$, then it would produce a detectable apocenter shift 
$\Delta \phi \ga 10\textrm{ mas}$ (see also \S~3.2 in Weinberg et al. 2005). 
Hall \& Gondolo (2006) fit the total measured mass concentration $M(<r)$ 
given in Ghez et al. (2005) assuming a power-law density profile and obtain 
an upper bound of $\approx 10^5 M_\odot$ between $0.001-1\textrm{ pc}$. 

The surface brightness of stars as a function of projected radius from 
Sgr A$^\ast$ is well measured down to a radius of $\sim$0\farcs 5 
($\sim$0.02 pc).
With an assumed constant mass to light ratio, the inferred stellar 
mass distribution between this inner radius and an outer radius of 10$\arcsec$ 
is consistent with
\begin{equation}
M_{\ast}(<r) = (6\times10^5 M_\odot) \left(\frac{r}{0.4\textrm{ pc}}\right)^{1.6}
\end{equation}
(Genzel et al. 2003b; see also Sch\"odel et al. 2007).  Extrapolating this profile down to a radius 
of $0.01\textrm{ pc}$ gives an enclosed mass 
$M_{\ast}(<0.01\textrm{ pc}) \approx1-2\times10^3M_\odot$. Furthermore, theoretical 
estimates of the density of cold dark matter halo particles suggest that $\sim 1000 M_\odot$ of dark 
matter might reside in the inner $0.01 \textrm{ pc}$ of the GC 
(Gondolo \& Silk 1999; Ullio et al. 2001; Merritt et al. 2002, Gnedin \& Primack 2004). 
Likewise, the mass contribution from a cluster of stellar remnants, as 
predicted by Morris (1993) and Miralda-Escud\'e \& Gould (2000),
is expected to be $\sim 1000 M_{\odot}$ within 0.01 pc.  
Unfortunately, these
estimates are all smaller than the current upper-bound by a factor of 
$\approx100$.  Measurements of stellar orbits with a next generation large 
telescope are, however, expected to be sensitive to an extended mass 
distribution of magnitude $M_{\ast}(<0.01\textrm{ pc}) \approx10^3M_\odot$ 
(Weinberg et al. 2005).

\section{DISCUSSION}
\label{sec:disc}

Orbit modeling of astrometric and radial velocity measurements 
of short period stars provides a direct estimate of the Milky Way's central
black hole mass and distance.  Our analysis of S0-2's orbit yeilds 
a black hole mass of 
M$_{bh}$ = 4.1 $\pm$ 0.6 $\times$ 10$^6  M_{\odot}$ and 
distance of R$_0$ = 8.0 $\pm$ 0.6 kpc, 
if nothing is assumed about
the black hole's intrinsic motion.  If we assume that the black hole
has no intrinsic motion relative to the central stellar cluster
(i.e., no massive companion),
but still allow for systematics in the reference frames,
then we obtain M$_{bh}$ = 4.5 $\pm$ 0.4 $\times$ 10$^6  M_{\odot}$ and 
distance of R$_0$ = 8.4 $\pm$ 0.4 kpc. 
This study shows that there are three systematic errors that must
be accounted for to obtain accuracy in estimates of orbital parameters and
this leads to larger uncertainties than have been reported in the past.
First, since a dominant source of systematic error in the data set appears to
be source confusion (see \S3 \& 4), we use only data from the brightest short 
orbital period star, S0-2, and only those measurements that are not 
confused with other known sources.  Second, the motion of the 
black hole relative to the measurements' reference frame should be left as a 
free parameter, to account for both any possible intrinsic motion of 
the black hole as well as systematics in the astrometric or spectroscopic 
reference frames.  
Third, while 
SgrA*-IR is detected with a precise position in deep LGSAO images, it appears 
to be biased; therefore, the position of the black hole should be treated as a 
free parameter in the fits in spite of the temptation to reduce the degrees of 
freedom with this detection.  Because these systematics were not 
incoporated into earlier simultaneous estimates of
M$_{bh}$ and R$_0$ from the orbit of S0-2, the uncertainties
in these initial studies were significantly underestimated;
Eisenhauer et al. (2003, 2005), who do not account for the first two 
systematics, obtain M$_{bh}$ = $3.6 \pm 0.3 \times 10^6 M_{\odot}$ and R$_0$ = 
7.6 $\pm$ 0.3 kpc.  Ghez et al. (2005a) used S0-2, S0-16, and S0-19 
simultaneously, and allowed $V_x$ and $V_y$ to be free parameters, 
to derive a mass at fixed R$_o$ 
of 3.7 $\pm$ 0.2 $\times$ 10$^6$ (R$_0$/ 8 kpc)$^3$ M$_{\odot}$, which 
was pulled down
by the two astrometrically-biased fainter stars, while Ghez et al. (2003) 
obtained a mass estimate of
4.1 $\pm$ 0.6 $\times$ 10$^6$ (R$_0$/ 8 kpc)$^3$ M$_{\odot}$ from S0-2
alone.
If we ignore the first two effects in model fits to our data, as was done
by Eisenhauer et al. (2003, 2005; the only other work to estimate R$_0$ from orbits),
we obtain a poor quality fit ($\chi^2/dof$ = 2.0), uncertainties that are a factor of 2 smaller, and somewhat higher values than what we report in 
Table \ref{tbl_orb} ($M_{bh}$ = $4.7 \pm 0.3 \times 10^6 M_{\odot}$ and 
R$_0$ = 8.6 $\pm$ 0.2 kpc).  
The removal of biased astrometric points 
dominates the shift in the black hole's mass and distance to lower values in 
our analysis.  This is somewhat suprising as this would suggest that similar 
removal of biased points might lower the Eisenhauer et al. (2005) results.  
However the biases may differ, as their early astrometric data 
measurements were made at three times lower angular resolution.  
An astrometric reference frame drift could also 
explain this effect, since $V_x$ and $V_y$ were held fixed in their analysis.
The addition of $V_z$ as a free parameter dominates the resulting uncertainties.
In summary, in order to get an accurate measure of M$_{bh}$ and R$_0$ from
modeling of the short period orbits at the Galactic center, 
it is critical to account for the three sources of systematics 
described above.


The black hole mass measured here from a stellar orbit is larger than the 
$\sim 2-3 \times 10^6 M_{\odot}$ inferred from 
using projected mass estimators, which rely on 
measured velocity dispersions 
(e.g, Eckart \& Genzel 1997; Genzel et al 1997; Ghez et al. 1998;
Genzel et al. 2000; see also Chakrabarty \& Saha 2001).  
This difference most likely arises from
the assumptions intrinsic to the use of projected mass estimators.
In particular, 
the projected mass estimators are based on the assumption that
the entire stellar cluster is measured, which is not the case
for the early proper motion studies as their fields of view were quite 
small (r $\sim$ 0.1 pc).  Such pencil beam measurements 
can lead to significant biases (see discussions in Haller et al. 1996; 
Figer et al. 2003).
An additional bias can arise if there is a central depression in the stellar 
distribution, such as that suggested by
Figer et al. (2003).  These biases can introduce factors of 2 
uncertainties in the values of the enclosed mass obtained from
projected mass estimates and thereby
account for the difference between
the indirect mass estimate from the velocity dispersions and the
direct mass estimate from the orbital model fit to S0-2's kinematic data.

A higher mass for the central black hole brings our Galaxy into 
better agreement with the $M_{bh} - \sigma$ relation
observed for nearby galaxies (e.g., Ferrarese  \& Merritt 2000; 
Gebhardt et al. 2000; Tremaine et al. 2002).   
For a bulge velocity dispersion that corresponds to that of
the Milky Way ($\sim$103 km s$^{-1}$; Tremaine et al. 2002),
the $M_{bh}-\sigma$ relationship from Tremaine et al. (2002) 
predicts a black hole mass of 9.4 $\times$ 10$^6$ M$_{\odot}$,
which
is a factor of 5 larger than the value of the Milky Way's
black hole mass used by these authors (1.8 $\times$ 10$^6$ M$_{\odot}$ from
Chakrabarty \& Saha 2001).  
The black hole mass presented here 
of 4.1 $\pm$ 0.6 $\times$ 10$^6 M_{\odot}$
brings the Milky Way more in line with this
relationship.  With one of the most accurate and lowest central black hole 
masses, the Milky Way is, in principle, an important anchor for
the $M_{bh}-\sigma$ relationship.  However, the velocity dispersion of the 
Milky Way is much more uncertain than that of other nearby galaxies. 
Therefore our revised mass has only modest impact
on the coefficients of the $M_{bh}-\sigma$ relation.

Revision of the central black hole's mass and distance can also, in principle,
 impact our understanding of the structure within our galaxy 
both on small and large scales.
On the large scale, if we assume that the black hole is located
at the center of our Galaxy, then its distance provides a measure
of $R_0$.  Its value from this study is consistent with the 
IAU recommended value of 8.5 kpc as well as the value of 8.0 $\pm$ 0.5 kpc 
suggested by Reid (1993), based on a ``weighted average"\footnote{consensus 
value with consensus errors} of all prior indirect
measurements of $R_0$.  Combining the value for $R_0$ from this study 
with the proper motion of Sgr A* along the direction of Galactic longitude measured with VLBA in the radio quasar reference frame 
(Reid \& Brunthaler 2004; $\mu_{SgrA*,
long}$ = -6.379 $\pm$ 0.026 
mas yr$^{-1}$) and the Sun's deviation from a circular orbit (Cox 2000; 
12 km s$^{-1}$) in the direction of Galactic rotation, we obtain an estimate 
of the local rotation speed, $\theta_0$, of 229 $\pm$ 18 km s$^{-1}$, which 
is statistically consistent with other measurements; 
these include a value of 222 $\pm$ 20 km s$^{-1}$ from the review of 
Kerr \& Lynden-Bell (1986) and 270 km s$^{-1}$ derived by M\'endez et al. (1999) 
from the absolute proper 
motions of $\sim$30,000 stars in the 
Southern Proper-Motion survey. 
As two of the fundamental Galactic constants, 
$R_0$ and $\theta_0$ are critical parameters for
axisymmetric models of the Milky Way.  Under the assumption that the
stellar and gas kinematics within our Galaxy are well measured, the values
of $R_0$ and $\theta_0$ determine the mass and shape of the Milky Way
(Olling \& Merrifield 2000; Olling \& Merrifield 2001).  
Of particular interest is
the value of the short-to-long axis ratio of the dark matter halo, q,
as it offers a valuable opportunity to distinguish between 
different cosmological models.
As Olling \& Merrifield (2001) demonstrate, 
the uncertainty in q for
the Milky Way is dominated by the large uncertainties in $R_0$ and
$\theta_0$.  While our uncertainties in R$_0$ are currently too large to 
constrain q, future precision measurements of R$_0$ through stellar orbits
may be able to do so and could thereby possibly distinguigh between 
various dark matter candidates (Olling \& Merrifield 2001).

Closer to the black hole, knowing its mass and distance from the Sun
improves our ability to study the kinematics of stars within its
sphere of influence.  
Much less kinematic information is needed to determine the orbital parameters
for stars whose motion is dominated by the gravitational influence
of the central black hole; for instance, with only measurements of a star's position, velocity, and
accelaration in the plane of the sky along with a single line of sight 
velocity, a complete orbital solution can be derived once the black hole's
mass and distance are well contrained.
Improved constraints on
the central black hole's properties and their degeneracies, as presented
here, along with improved astrometry, has allowed us to derive orbital 
information for individual stars at much larger galacto-centric distances.
With these measurements, in Lu et al. (2006, 2008), we test for the existence 
and properties of the young stellar disk(s), proposed by
Levin \& Beloborodov (2003) and Genzel et al. (2003b) from a statistical
analysis of velocities alone.  The direct use of individual stellar orbits 
out beyond a radius of 1$\tt''$ reveals only 
one, relatively thin, disk of young stars (Lu et al. 2008).   

On an even smaller scale, the mass and distance of the black hole set the
magnitude and time-scale for various relativistic effects.  
Given estimated Keplerian orbital elements for stars at the Galactic center, 
we expect to be able to measure their stellar orbits with sufficient
precision in upcoming years to detect the Roemer time delay, the
special relativistic transverse Doppler shift, the general relativistic
gravitational red-shift, and the prograde motion of periapse
(e.g., Weinberg et al. 2005; Zucker \& Alexander 2007).  
These effects will most likely be measured with S0-2 first, as  
it has the shortest orbital period (P=15 yr), is quite eccentric (e=0.89)
and, as one of the brighter stars ($K_{S0-2}$ = 14 mag), it can be measured 
with the greatest astrometric and spectroscopic accuracy. 
The radial velocity signatures of the first three effects are expected to 
be comparable to each other and will impart a $\sim$200 km/s deviation at
closest approach (Zucker \& Alexander 2007), when the star is predicted to have a 
line of sight velocity of -2500 km/s based on our updated Keplerian model.  
This effect is large compared to the radial velocity precision ($\sim$ 20 
km/sec).  Likewise, the expected apoapse center shift for S0-2, 
$\Delta s = \frac{6 \pi G M_{bh}}{R_0 (1-e) c^2} = 0.9$ mas
(see e.g., Weinberg 1972; Weinberg et al. 2005),
is an order of magnitude larger than our current measurement 
precision ($\sigma_{pos}$ $\sim$ 0.1 mas).  
Improved adaptive
optics systems on existing telescopes and larger telescopes
(see Weinberg et al. 2005) will improve the sensitivity to the predicted
apocenter shift.
To put this measurement into context with 
existing tests of general relativity, it is useful to note that one of the 
strongest 
constraints on general relativity to date comes from
the Hulse-Taylor binary pulsar, PSR 1913+16, which has a
relativistic parameter at periapse, $\Gamma = r_{sch} / r_{periapse}$, of 
only $5 \times 10^{-6}$, $\simeq$ 3 orders of magnitude smaller than that
of S0-2 (Taylor \& Weisberg 1989; Zucker \& Alexander 2007).  
The stars at the Galactic center are therefore probing
an unexplored regime of gravity in terms of the relativistic object's mass
scale and compactness.

Precession from general relativistic effects also influences the
timescale for resonant relaxation processes close to the black hole
(see, e.g, Rauch \& Tremaine 1996; Hopman \& Alexander 2006).
When precession from general relativity dominates over that from the extended
mass distribution, the resonant relaxation timescale is proportional
to $M_{bh}^2 \times (J_{LSO}/J)^2 \times P$, where $J$ and $J_{LSO}$ are the 
orbital angular momenta for the orbit of interest and at the last stable 
circular 
orbit around the black hole, respectively, and P is the orbital period.
For a given semi-major 
axis and accounting for the linear mass dependence of $(J_{LSO}/J)^2$,
this results in a $M_{bh}^{5/2}$ dependency.  Thus the higher black hole mass inferred from this study
increases the timescale over which the black hole's loss cone would be 
replenished in the regime where general relativity dominates.  
For the regime where the
extended mass distribution dominates, the resonant relaxation timescale
scales only as $M_{bh}^{1/2}$.
A higher black hole mass 
also implies a longer period for the innermost stable circular orbit.  If the central black 
hole is non-spinning, the innermost stable circular orbit has a period of 31 $\frac{M_{bh}}{4.1 \times 10^6 M_{\odot}}$ min.  Periodicities on shorter 
timescales, such as the putative quasi-periodic oscillation (QPO) at 
$\sim$ 20 min (Genzel. et al.
2003a; Eckart et al. 2006; B\'elanger et al. 2006) 
have been interpreted as arising from the innermost stable circular orbit
of a spinning black hole.  At the present mass, the spin would have 
to be 0.6 of its maximal rate to be consistent with the possible periodicity.  
However, it is important to caution that other mechanisms can give rise
to such short periodicities, such as a standing wave pattern recently
suggested by Tagger \& Melia (2006).   
Furthermore, claims of a QPO in SgrA* have been called into question; Do et al. 
(2008) find that the near-IR temporal power spectrum of SgrA* is statistically 
consistent with pure red noise, such as might be caused by disk instabilities 
or intermittent jet fluctuations, and Belanger et al. (in preparation) reach a 
similar conclusion for the X-rays variations.
 
\section{CONCLUSIONS}

The short orbital period star S0-2 has been intensively studied
astrometrically (1995-2007) and 
spectroscopically (2000- 2007) with the W. M. Keck 10 meter telescopes.
Fits of a Keplerian orbit model to these data sets, after removing
data adversely affected by source confusion, result in estimates
of the  black hole's mass and distance of  
$4.1 \pm 0.6 \times 10^6  M_{\odot}$ and 8.0 $\pm$ 0.6 kpc, respectively.
While the current analysis is dominated by 11 years of astrometric 
measurements that have $\sim$ 1.2 mas uncertainties, the LGSAO over the 
last 3 years have positional uncertainties that are an order of magnitude 
smaller (100-200 $\mu$as).   With higher strehl ratios and more sensitivity, 
LGSAO measurements
are also less affected by source confusion; this is especially important
for the closest approach measurements, which have to contend with 
source confusion from the variable source SgrA*-IR.  
Following S0-2 for another 10 years should result in the measurement of
the Sun's peculiar motion in the direction of the Galactic center from
the orbit of S0-2 with a precision of a few km s$^{-1}$ and 1\% measurement 
of $R_0$.  
At this precision, the measurement of $R_0$ is of particular interest 
because it could reduce the uncertainty in the cosmic distance ladder.

\acknowledgements
We thank the staff of the Keck observatory, especially
Joel Aycock, Randy Campbell, Al Conrad, Jim Lyke, David LeMignant,
Chuck Sorensen, Marcos Van Dam, Peter Wizinowich, and director Taft 
Armandroff, for all their help in obtaining the new 
observations.  We also thank Brad Hanson, Leo Meyer, and Clovis Hopmann for 
their constructive comments on the manuscript, and the referee, Rainer 
Schodel, for his helpful suggestions.
Support for this work was provided by NSF
grant AST-0406816 and the NSF Science
\& Technology Center for AO, managed by UCSC
(AST-9876783), and the Levine-Leichtman Family Foundation.
The W. M. Keck Observatory,
is operated as a scientific partnership among the California Institute
of Technology, the University of California and the National Aeronautics and
Space Administration.  The Observatory was made possible by the generous
financial
support of the W. M. Keck Foundation.  The authors wish to recognize and
acknowledge the very significant cultural role and reverence that the summit
of Mauna Kea has always had within the indigenous Hawaiian community.  We are
most fortunate to have the opportunity to conduct observations from this
mountain.


\appendix

\section{Cluster Reference Frame}
\label{app:cluster_ref_frame}

All positional measurements from the individual images ($X', Y'$) 
are transformed with a full first order polynomial to 
a common reference system ($X, Y$), which we refer to as the cluster reference
frame (see Ghez et al. 1998, 2000, 2005a; Lu et al. 2008).  
The transformations are derived 
by minimizing the net displacements, allowing for proper motions, of all
the coordinate reference stars (see \S3.1) relative to their positions 
in a common reference image ($ref$), 
which for this study is the 2004 Jul LSGAO image.
Specifically, we minimize
the following sum over the coordinate reference stars ({\it s}): 
\[  D = \displaystyle\sum_{s}^{N_{stars}} (\Delta X_{s, e}^2 + \Delta Y_{s,e}^2) / W_{s,e}, \]
where 
$\Delta X_{s,e} = X_{s,ref} + V_{x_s} \times (t_e - t_{ref}) - X_{s,e}$,
$\Delta Y_{s,e} = Y_{s,ref} + V_{y_s} \times (t_e - t_{ref}) - Y_{s,e}$, and
$W_{s,e} = \sigma_{\Delta X'_{s,e}} + \sigma_{\Delta Y'_{s,e}}$, and 
where $X$ and $Y$ are expressed as the following function of the
measured positions $X'$ and $Y'$ for
for each epoch ($e$) 
\[ X_e = a_{0_e} + a_{1_e} \times X'_e + a_{2_e} \times Y'_e, \]
\[ Y_e = b_{0_e} + b_{1_e} \times X'_e + b_{2_e} \times Y'_e. \]
The coefficients for the reference epoch are fixed to 
$a_{1_{ref}} = b_{2_{ref}} = 1$ and 
$a_{0_{ref}} = a_{2_{ref}} = b_{0_{ref}} = b_{1_{ref}} = 0$ and 
the coefficients for the remaining epochs 
($a_{0_e}$, $a_{1_e}$, $a_{2_e}$, $b_{0_e}$, $b_{1_e}$, and $b_{2_e}$)
come from the minimization of $D$.
Because of degeneracies between
coordinate transformations and proper motions of the coordinate
reference stars, the net displacment is minimized in two steps.
First, $D$ is minimized with the proper motions ($V_x$ and $V_y$) 
of the coordinate reference stars set to zero
in order to obtain preliminary transformation coefficients.
Using these initial coeffficients, 
we transform  all the positional 
measurements to a common coordinate system
and fit a linear motion model to them in order to  
derive a first pass estimate of the proper motions.
Second, $D$ is minimized again, using the preliminary proper motions and
holding them fixed, while the final transformation coefficients are derived. 
This procedure produces proper motions for the 
coordinate reference stars that have no significant mean motion.
We therefore conclude that the resulting cluster reference frame is stable 
and free of significant systematics.

\begin{figure*}
\plottwo{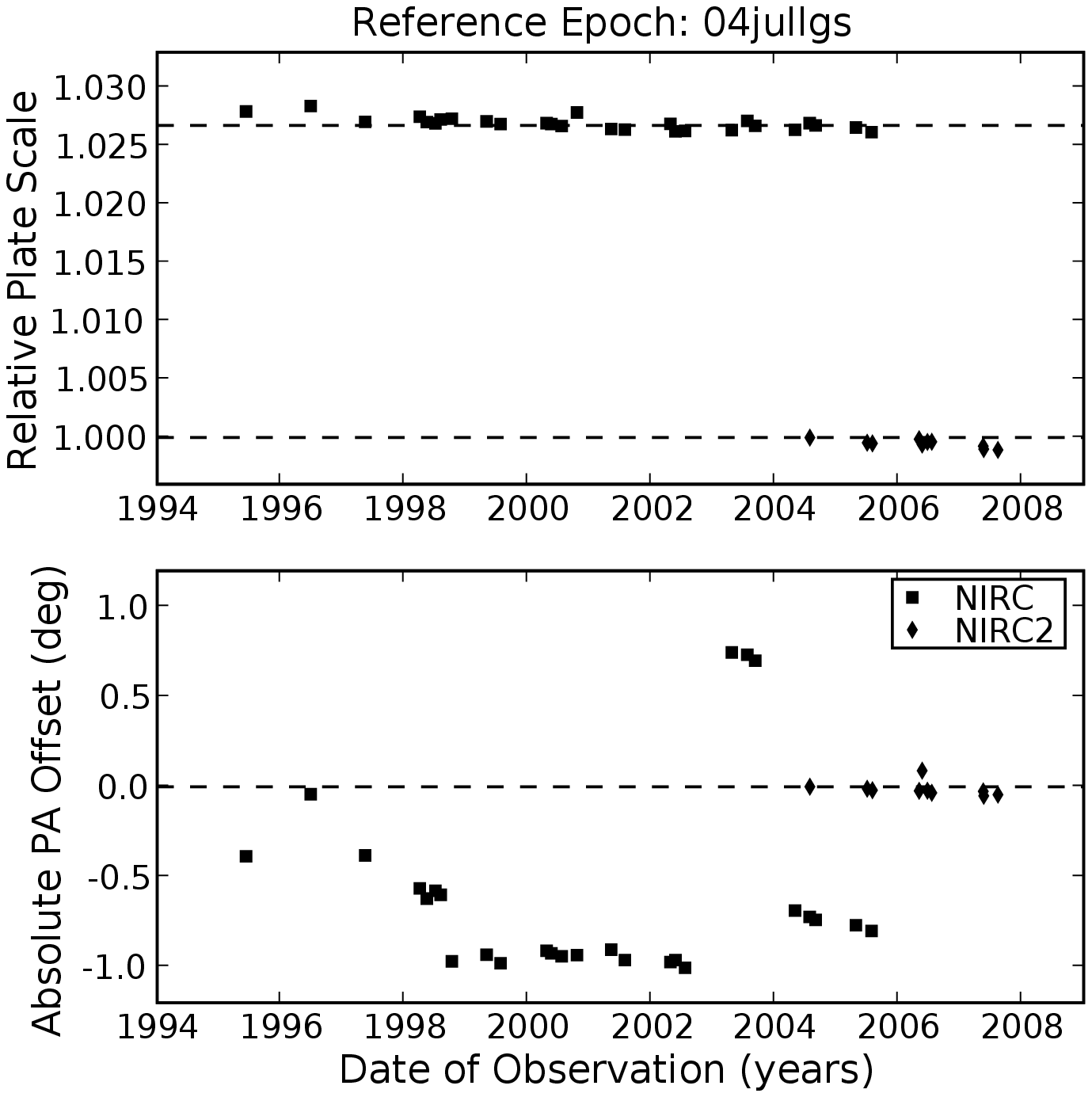}{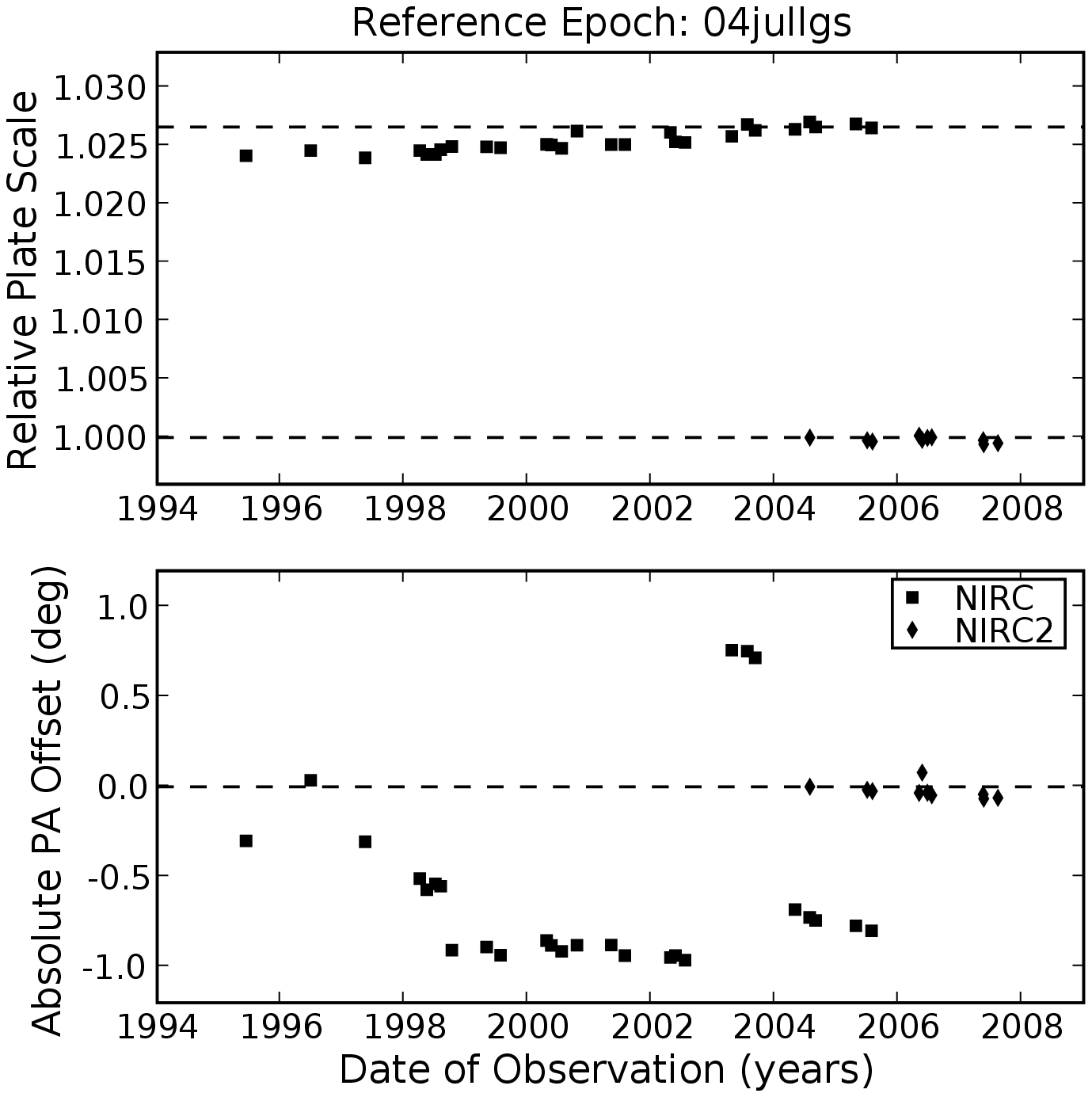}
\figcaption{
The plate scale {\it (top)} and the position angle {\it (bottom)}
over time for all data sets aligned using a set of stars that
excludes {\it (left)} and includes {\it (right)} the known young stars. 
The plate scale is relative to the
plate scale in the reference epoch of 2004 July LGS. The position
angle is the absolute position angle offset from the value reported
in the NIRC and NIRC2 instrument headers. 
Once the young stars are excluded, the estimated 
plate scales for NIRC ({\it squares}) and NIRC2 ({\it diamonds})
are very stable, approximately 0.05\% and 0.03\% (rms), respectively,
over multiple years.
NIRC shows several systematic jumps in the position angle relative to 
the value reported in the image headers, which is most likely a result
of instrument or telescope changes. 
\label{trans}
}
\end{figure*}

This procedure is also used to check the stability of the combined
effects of the camera systems and the coordinate reference stars.
By carrying out transformations that allow for only translation,
rotation, and a scale change, we examine the apparent stability 
of the camera's pixel scale and angle relative to that recorded 
in the header.   Figure \ref{trans} shows, on the left-hand side, 
that the relative pixel
scales for the cameras are stable to within 0.053\% (rms)
over the time baseline of this study and that the uncertainty in the
angle relative to the header PA is dominated by inaccuracies in the
header value (most of the jumps correspond to times when the camera
is known to have been opened for engineering purposes).  This also
provides a measure of the resampled NIRC pixel scale relative to the NIRC2 
pixel scale (1.0269 $\pm$ 0.0005) and an absolute NIRC pixel scale 
of 20.46 $\pm$ 0.01 mas pix$^{-1}$ when combined with the absolute
NIRC2 pixel scale from Appendix C.
On the right-hand side, Figure \ref{trans} displays the results
of the same excerise but using a set of coordinate reference stars
that includes the known young stars; the clear systematic
trend in the relative pixel scales demonstrates
the importance of removing this set of stars with known net rotation
from the coordinate reference star list. 

\section{NIRC2 Geometric Optical Distortions}
\label{app:nirc2_distortion}

Relative stellar positions from the deep LGSAO images have accuracies 
($\sim$0.2 mas) that are an order of magnitude smaller than the currently 
available optical distortion map for NIRC2 
(\url{http://www2.keck.hawaii.edu/inst/nirc2/preship\_testing.pdf}).
Since LGSAO/NIRC2 data was obtained with four different 
setups (e.g., centerings and/or position angles on the sky),
imperfections in the optical distortion corrections can introduce 1-2 mas
systematics, if unaccounted for, into the relative positions of
S0-2 (and S0-16).
We therefore introduce two steps into our analysis to correct for this effect.
First, we add, in quadrature, an additional 0.88 mas 
to all the LGSAO positional measurements
of the coordinate reference stars, such that the proper motions and
hence coordinate transformations are not biased.  
The magnitude of this term is derived by finding the value that reduced
the average offset of these LGSAO points from the linear proper motion
fits, which exclude these points, from 5$\sigma$ to 1$\sigma$.
Second, we derive explicit correction terms for the local optical distortions
for S0-2 and S0-16 positions in each of LGSAO epochs not obtained with
the same set up as the reference image (2004 July), using 
the orbits of 5 ``calibration" stars (S0-3, S0-7, S0-19, S0-26, and S0-27) 
that are within 0\farcs 5 of S0-2. 
These terms are obtained by first using only the speckle data, 
which are distortion calibrated with respect to the reference 
image (2004 July/LGSAO; see Lu et al. 2008), 
the reference image (taken with setup\#1), and the 
one other LGSAO image taken with the same setup as the reference image to solve for the orbits of the 5 calibration stars.  
For each LGSAO epoch not included in these fits, the average
offsets of these five stars' aligned measurements from their predicted 
location is used to characterize the residual distortions for 
that image (relative to the reference image) at the position of S0-2 and S0-16
and the standard deviation of the offsets provide an estimate of the  
uncertainties in these values.  Setup \#3 is the only LGSAO observational 
configuration, other than that used for the reference image, used in
multiple epochs.  From the measurements with setup \#3, it can be seen that 
the rms of their estimated bias terms (0.24 mas) is smaller than 
the uncertainty in each bias term estimated
from the rms of the 5 stars ($\sim$ 0.67 mas).
This suggests that the bias terms are relatively static (see also Appendix
A) and that their
uncertainties are dominated by our uncertainties in the stellar orbits
(and possible structure in the distortion on scales $<$0\farcs 5).  
We therefore derive an average bias correction value and uncertainty 
for each setup.  The final bias terms, which range in value between 
1.6 and 2.6 mas, are added to the LGSAO positional measurements made 
with setups \#2-4 in the 
analysis presented in \S\ref{sec:orbit} and their uncertainties are
added in quadrature with the uncertainties associated with centroiding and
coordinate transformation; this bias term as already been incorporated into the
values and uncertainties reported in Table \ref{tbl_pos}.  Correlations in the 
bias corrections for setup \#3 are applied and accounted for in the Monte 
Carlo simulations described in \S4.  

\section{Absolute Astrometry}
\label{app:absolute_astrometry}

\begin{figure*}
\epsscale{1.0}
\plotone{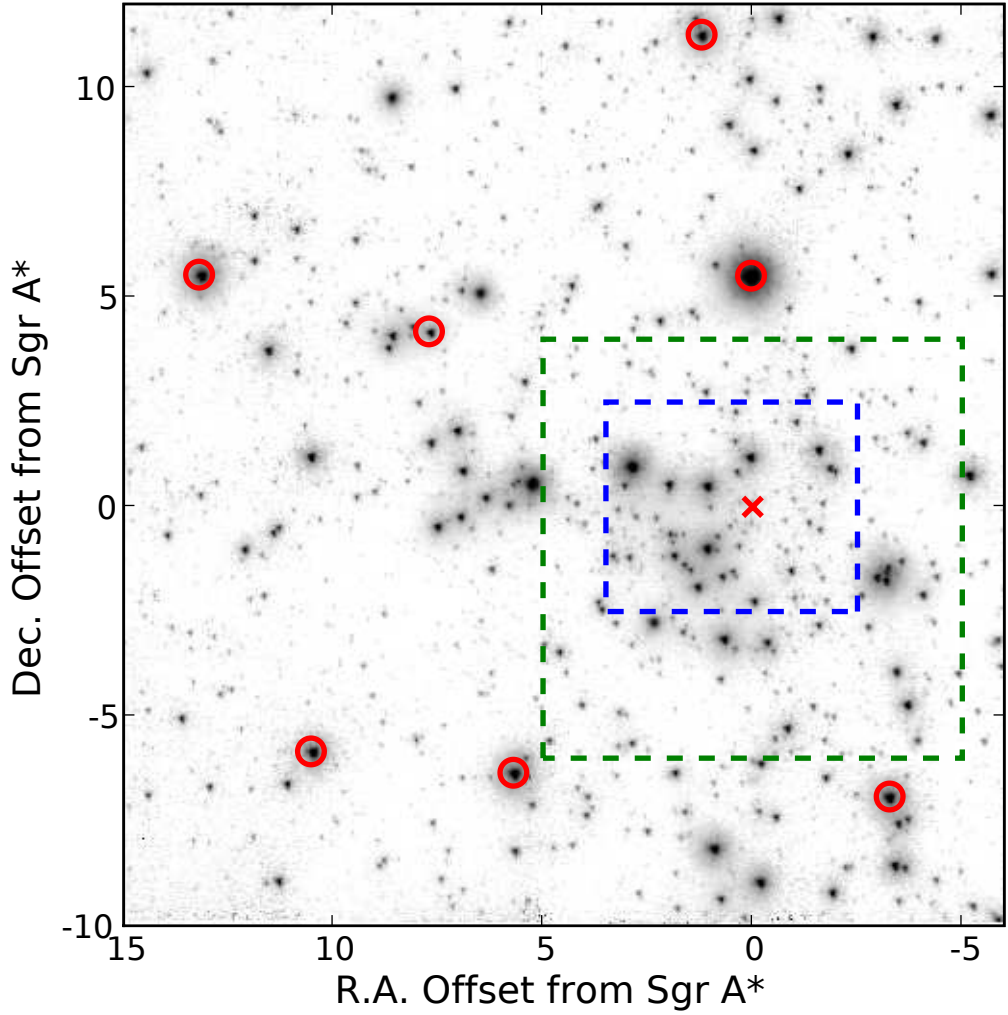}
\figcaption{
Infrared mosaic measuring the positions of the SiO masers.  The 7 masers,
whose radio positions are well measured by Reid et al. 2007 and which
used to establish an absolute reference frame, are circled.  Dotted
lines depict the outline of example LGSAO (green) and speckle (blue) 
images in which the short period stars are measured and placed in
the cluster reference frame.  
Since the masers sparsely sample the area of interest, only low order
polynomials are used to calibrate the cluster reference frame (i.e., pixel 
scale, orientation, and position of SgrA*-Radio).  
\label{fig:masers}
}
\end{figure*}

An absolute astrometric reference frame for the Galactic center
was established from radio observations of seven SiO masers 
(Reid et al. 2003, 2007). Relative measurements in the infrared were tied
to the absolute frame by observing, in the infrared, the red-giant stars
that are the source of the maser emission (Figure \ref{fig:masers} \& Table \ref{tab:masers}).
Observations were taken in 2005 June, 2006 May, and 2007 August
using LGSAO/NIRC2 (see \S\ref{sec:obs}) with
$10.86 ~ \textrm{s}$ integrations in the K' band, each composed of 60 co-added 0.181 s
exposures in order to avoid saturating the bright masers.
A nine position dither box pattern was used to construct a
22$\arcsec\times$22$\arcsec$ mosaic with two exposures at each position
for the 2005 mosaic and three exposures at each position for the 2006 and 2007
mosaics.
The individual frames for each data set were cleaned, undistorted, and then
registered and mosaicked using the IRAF {\it xregister} and {\it drizzle}
routines.
Subset-mosaics were also created with only 1 exposure at each position and
were used to derive centroiding uncertainties.
StarFinder was run on the resulting mosaicked images to extract stellar
positions and uncertainties from the RMS error of the subset-mosaics.
Centroiding errors were typically on the order of 1.4 mas.
This yields an IR starlist for each epoch with positions in NIRC2 pixel
coordinates.

\begin{deluxetable}{lrrrrrrr|r}
\tabletypesize{\tiny}
\tablewidth{0pt}
\tablecaption{Maser Properites}

\tablehead{
  \colhead{} &
  \colhead{IRS 9} &
  \colhead{IRS 7} &
  \colhead{IRS 12N} &
  \colhead{IRS 28} &
  \colhead{IRS 10EE} &
  \colhead{IRS 15NE} &
  \colhead{IRS 17} &
  \colhead{Average} 
}
\startdata
K Magnitude &  9.1 &  7.7 &  9.5 &  9.3 & 11.3 & 10.2 &  8.9 & - \\ 
X Position (arcsec) &   5.679 &   0.032 &  -3.264 &  10.484 &   7.684 &   1.209 &  13.139 & - \\ 
Y Position (arcsec) &  -6.332 &   5.529 &  -6.912 &  -5.833 &   4.196 &  11.268 &   5.560 & - \\ 
X Velocity (mas/yr) &   3.06 &  -0.58 &  -1.06 &   2.00 &   0.04 &  -1.96 &  -1.61 & - \\ 
Y Velocity (mas/yr) &   2.11 &  -3.52 &  -2.70 &  -5.29 &  -2.09 &  -5.68 &  -0.75 & - \\ 
\tableline
$[$IR - Radio$]$ X Position (mas) &  3.7 $\pm$  5.2 &  3.5 $\pm$  7.2 & 11.6 $\pm$  7.3 &  0.6 $\pm$  5.9 & -3.9 $\pm$  3.4 & -1.9 $\pm$  5.5 & -5.3 $\pm$  5.2 &  1.2 $\pm$  5.7 \\ 
$[$IR - Radio$]$ Y Position (mas) & -4.1 $\pm$  4.8 & -3.4 $\pm$  7.4 &  4.4 $\pm$  5.4 & -4.5 $\pm$  5.9 &  0.9 $\pm$  5.0 &  9.4 $\pm$  7.5 & -6.4 $\pm$  6.7 & -0.5 $\pm$  5.7 \\ 
$[$IR - Radio$]$ X Velocity (mas/yr) &  1.4 $\pm$  0.8 &  0.8 $\pm$  0.5 &  1.7 $\pm$  1.0 & -0.3 $\pm$  2.6 & -0.1 $\pm$  0.3 & -1.0 $\pm$  0.4 &  1.7 $\pm$  1.2 &  0.6 $\pm$  1.1 \\ 
$[$IR - Radio$]$ Y Velocity (mas/yr) &  0.4 $\pm$  0.8 & -1.7 $\pm$  0.6 &  0.7 $\pm$  1.0 & -0.5 $\pm$  2.6 & -1.5 $\pm$  0.3 &  0.0 $\pm$  0.4 & -3.5 $\pm$  1.3 & -0.9 $\pm$  1.5 \\ 
\enddata 
\tablecomments{Maser positional uncertainties and differences
are averaged over the 3 maser epochs: 2005.495, 2006.336, 2007.612.  The values
in the last column are the average and standard deviation of the values for
the individual masers.} 
\label{tab:masers}
\end{deluxetable}

The radio maser positions were propagated forward using
velocities from Reid et al. 2007 to create a radio maser starlist
at the epoch of each of the above IR mosaics.  Uncertainties in these propogated
radio positions are, on average, $\sim$1.4 mas.
For each epoch, the IR maser starlist was aligned to the Radio mosaic starlist,
which resulted in a new IR mosaic starlist in the absolute astrometric
reference frame with Sgr A*-Radio at the origin.
This alignment process used only four independent parameters
(a global pixel scale, a rotation, and an origin in the x and y directions)
to transform between the NIRC2 coordinate system of the IR mosaics to
the absolute coordinate system of the radio masers. 
While using higher order polynomial 
transformations reduce the residual offsets positions from SgrA* between
the infrared and radio measurements,
we conservatively chose to use this low order 
transformation to capture within the uncertainties the possible impact of 
systematics, such as
uncorrected residual camera distortions and differential atmospheric refraction.
This is particularly important given the sparse sampling of masers across
the rgion of interest (see Figure \ref{fig:masers}).
Uncertainties in the transformation to absolute coordinates, which
were determined with a half-sample bootstrap Monte Carlo simulation of
100 iterations where each iteration uses only half the stars in each starlist,
were added in quadrature to the infrared centroiding uncertainties to produce
a final uncertainites in the infrared absolute positions of the masers.
After the transformation to absolute coordinates, the absolute value of
the offsets between the positions of the masers relative to SgrA* measured 
in the infrared and radio are on average 0.8$\sigma$ and  
0.8 $\sigma$, or equivalenly,
5.7 mas and 5.7 mas in the x and y direction, respectively
(see Table \ref{tab:masers}); we take this
to be our uncertainty in the position of Sgr A*-Radio in the infrared 
maser mosaic.  Likewise, the transformations between the infrared and 
radio reference frame yields a plate scale of 9.963 $\pm$ 0.005
mas/pix and a position angle offset for NIRC2 of 0.13$^\circ \pm$ 0.02$^\circ$.
Each of the three infrared maser mosaics yields comparable results
(see Table \ref{tab:nirc2astrometry}). 
Uncertainties in the absolute positions in the  
infrared reference frame are dominated by residual 
optical distortions, which are amplified by the large dithers 
necessary to construct the mosaics.  

\begin{deluxetable}{lrrr}
\tabletypesize{\scriptsize}
\tablewidth{0pt}
\tablecaption{NIRC2 Absolute Astrometry}

\tablehead{
  \colhead{} & 
  \colhead{2005.495} & 
  \colhead{2006.336} & 
  \colhead{2007.612}  
}
\startdata
Plate Scale (mas/pixel) & 9.963 $\pm$ 0.005 & 9.964 $\pm$ 0.004 & 9.961 $\pm$ 0.006  \\ 
Angle (deg) &  0.12 $\pm$  0.02 &  0.13 $\pm$  0.02 &  0.14 $\pm$  0.02  \\ 
\enddata
\tablecomments{The final values for the NIRC2 plate scale, 
the position angle of the NIRC2 columns with respect to North 
Sgr A* are averages over these values and their errors.} 
\label{tab:nirc2astrometry}
\end{deluxetable}

A comparison of the maser's proper motions as measured in the 
radio and the infrared provides an estimate of how accurately we 
can transform our relative measurements into a reference frame in which
SgrA*-Radio is at rest and the orientation is set by background quasars
(Reid et al. 2007).  The absolute infrared proper motions of the masers, as well
as all other stars detected in the infrared maser mosaics,
were derived by fitting a linear model to the positions as a function of time 
from the three IR maser starlists that were separately aligned to the radio 
reference frame.  Because the alignment uncertainties are dominated by
residual distortion and therefore correlated across epochs for a given maser, 
this source of uncertainty is not included in the linear proper motion 
modeling.  The differences in the proper motions measured
in the radio and in the infrared have an average value of 0.6 $\pm$ 0.4 mas/yr 
and -0.9 $\pm$ 0.6 mas/yr in the x and y directions, respectively, where the uncertainties are 
the standard deviation of the mean.  Therefore, at present, it is not possible 
to use these measurements to eliminate possible drifts in the cluster reference frame as the source of any apparent $V_x$ or $V_y$ from the orbital fits of S0-2
(see \S4.1).

The relative astrometry measurements presented in \S3 were transformed into this
absolute reference frame through a set of infrared stars we designated
as {\it infrared absolute astrometric standards}.  
Absolute astrometric standards were defined to be those stars
that are 1) detected in all three IR mosaics (2005, 2006, 2007),
2) outside the central arcsecond (r$>$0\farcs 5),
3) have velocities less than 15 mas/yr and velocity errors less than 5  mas/yr,
4) have reasonable velocity fits ($\chi^2/dof < 4$),
and 5) are brighter than K=15.  With absolute kinematics for 158 stars within 5'',
we solve for a 4 parameter transformation model by comparing the relative positions in the
reference epoch image, which are in instrumental pixel coordinates, 
 and the estimated absolute coordinates for that epoch, which are in arcsec relative to 
the position of SgrA*.  Since all other
epochs are aligned to this reference epoch, positional measurements
for all stars in all epochs are easily transformed into absolute coordinates.
While uncertainties in the absolute infrared reference frame 
dominate the final absolute positional uncertainties relative to SgrA*-Radio, 
they are a negligible source of uncertainty
for the orbital analysis.

\end{document}